\documentclass[aps,prl,twocolumn,superscriptaddress,longbibliography]{revtex4-2}
\usepackage{mathrsfs}
\usepackage{tabularx}
\usepackage{slashed}
\usepackage{amsmath}
\usepackage{amsfonts}
\usepackage{physics}
\usepackage{bm}
\usepackage{graphicx,color}
\usepackage{times}
\usepackage[caption=false]{subfig}
\usepackage[colorlinks,linkcolor=red,citecolor=blue]{hyperref}
\usepackage{float}

\usepackage{xcolor}
\definecolor{darkblue}{rgb}{0.,0.,0.4}
\definecolor{darkred}{rgb}{0.5,0.,0.}
\definecolor{BlueViolet}{RGB}{138,43,226}
\definecolor{SkyBlue}{RGB}{30,144,255}
\definecolor{DarkGreen}{RGB}{0,100,0}

\usepackage[normalem]{ulem}
\usepackage{comment}

\begin{document}
\title{
Robust non-Abelian  even-denominator fractional Chern insulator in twisted bilayer MoTe$_2$}

\author{Feng Chen}
\affiliation{Department of Physics and Astronomy, California State University Northridge, Northridge, California 91330, USA}
\author{Wei-Wei Luo}
\affiliation{Zhejiang Institute of Photoelectronics \& Zhejiang Institute for Advanced Light Source, Zhejiang Normal University, Jinhua, Zhejiang 321004, China}
\affiliation{School of Science, Westlake University, Hangzhou 310024, China, and \\
Institute of Natural Sciences, Westlake Institute of Advanced Study, Hangzhou 310024, China}
\author{Wei Zhu}
\email{zhuwei@westlake.edu.cn}
\affiliation{School of Science, Westlake University, Hangzhou 310024, China, and \\
Institute of Natural Sciences, Westlake Institute of Advanced Study, Hangzhou 310024, China}
\author{D. N. Sheng}
\email{donna.sheng1@csun.edu}
\affiliation{Department of Physics and Astronomy, California State University Northridge, Northridge, California 91330, USA}

\begin{abstract}
	\section{Abstract}
    A recent experiment observes a series of quantum spin Hall effects in transition metal dichalcogenide moir\'e  MoTe$_2$ [K. Kang, \textit{et. al}, Nature 628, 522-526 (2024)]. Among them, 
    the vanishing Hall signal at the filling factor $\nu=3$ implies a possible realization of a time-reversal pair of the even-denominator fractional Chern insulators (FCIs).  
    Inspired by this discovery, we investigate whether a robust incompressible quantum Hall liquid can be stabilized in the half-filled  Chern band of twisted MoTe$_2$ bilayers. We use the continuum model with parameters relevant to  twisted MoTe$_2$ bilayers and obtain three consecutive nearly flat Chern bands  with  the same  Chern number.
    Crucially, when the second moir\'e miniband is half-filled, 
    signatures of non-Abelian frctional quantum Hall  state are found via exact diagonalization
    calculations, including the stable six-fold ground state degeneracy that grows more robust with the lattice size and is consistent with  an even-denominator FCI  state.
    We further perform flux insertion simulations to reveal 
   a 1/2 quantized many-body Chern number as direct evidence of topological order. 
   Furthermore, the ground state density structure factors show no sharp peak, indicating no charge density wave order. 
   These evidences signal the potential of realizing the non-Abelian state at zero magnetic field in twisted bilayer MoTe$_2$ at the fractional hole filling 3/2.   
\end{abstract}
\maketitle

\section{Introduction}
Transition metal dichalcogenide (TMD) moir\'e systems have recently attracted great attention~\cite{cai2023signatures,park2023observation,zeng2023thermodynamic,xu_observation_2023} due to the discovery
 of emergent  fractional Chern insulators (FCIs)~\cite{sheng2011fractional,regnault2011fractional,neupert2011fractional, tang2011high,sun2011nearly} --- zero magnetic field analogues of the fractional quantum Hall (FQH) effect.  Theoretical studies~\cite{wu2019topological,li2021spontaneous,reddy2023fractional,crepel2023anomalous,morales2023magic,wang2023fractional,qiu2023interaction,Yu2023,jia2023moire,reddy2023toward,xwzhang2024} demonstrate the engineering of 
topologically nontrivial flat bands in moir\'e systems and the emergence of the FCI states such as
the Laughlin states and other Jain sequence states~\cite{reddy2023fractional, reddy2023toward}, which are  discovered experimentally~\cite{cai2023signatures,park2023observation,zeng2023thermodynamic,xu_observation_2023}. 
These results  imply the resemblance of the quantum geometry of the moir\'e Chern bands to that of the lowest Landau level~\cite{wang2021exact,Ozawa2021}. 
This immediately prompts an interesting question~\cite{fujimoto2024higher,reddy2024nonabelian,xu2024multiple,ahn2024landau,wang2024higher}: Is it possible to realize the first excited Landau level physics with non-Abelian statistics in the TMD moir\'e system?

More recently, Ref.~\cite{kang_observation_2024} reported a series of quantum spin Hall plateaus in mori\'e MoTe$_2$ system at hole fillings $\nu =2, 3, 4, 6$ with a twist angle around $\theta \approx 2.1^o$. Remarkably, this $\nu = 3$ state 
supports helical edge
modes with the half-quantized edge conductance $G =3/2 (e^2/h)$,  in sharp
contrast with the previously observed valley polarized states at odd integer fillings with $\theta\approx 3.7^o$ (e.g. $\nu=1$)~\cite{cai2023signatures,park2023observation}. 
Intuitively, a way to understand this $\nu = 3$ state is to examine possible phases in the second band~\cite{ChaomingJian2024,YahuiZhang2024, villadiego2024halperin}.
If the second moir\'e band is half-filled in each valley, it may produce a $\nu =1/2 +1/2$ fractional quantum spin Hall insulator in the valley-decoupled limit.
Together with the $\nu=2$ state from the filled lowest moir\'e band, it may provide an explanation for  the $\nu=3$ state observed experimentally~\cite{kang_observation_2024}.  
In the above scenario, one key point is if there is a fractionalized topological state at half filling of the second moir\'e band in each valley, 
which can survive the interaction between two valleys.

Besides the quantum spin Hall effect by filling a higher  moir\'e band, the spontaneous time reversal symmetry breaking promoted by Coulomb interaction makes the higher moir\'e band a promising platform for discovering distinct non-Abelian FCIs. 
In the study of the half-filled first excited Landau level, the leading topological ground state candidates include  the non-Abelian  
Pfaffian ~\cite{MOORE1991,Greiter1991}, anti-Pfaffian ~\cite{SSLee2007,Levin2007}, and particle-hole
symmetric Pfaffian states~\cite{Feldman2016}, while there are still intense debates on which one is realized in experiments~\cite{Heiblum2018,Mross2018,Simon2018,Simon2020}.  
A similarity between the second moir\'e band and first excited Landau level  has been illustrated~\cite{reddy2024nonabelian, fujimoto2024higher, ahn2024landau, xu2024multiple}  from both single particle information and many-body energy spectrum of the half-filled second moir\'e band
to indicate the possibility of realizing non-Abelian states.
However, the nature of the ground state 
for half filling of the second moir\'e band remains open. 
There are  competing states  including a charge density wave (CDW) state with no topological order, and other topologically ordered states distinctly different from the Landau level counterparts due to the absence of continuous translational symmetry~\cite{kol1993fractional,bernevig2012translation,sheng2024}.
These open issues motivate  the current work.

In this paper,  targeting the $\nu=3/2$  state  in moir\'e MoTe$_2$ for hole doping,  we address the physics of the second moir\'e band microscopically. 
Using the continuum model,  we demonstrate that the lowest three consecutive bands have Chern number $C=1$ by tuning model parameters between
available density-functional theory (DFT) values~\cite{reddy2023fractional,xwzhang2024,ahn2024landau}. 
At $\nu=3/2$,
 the ground state demonstrates a spontaneous ferromagnetism (FM) and valley polarization
 driven by Coulomb interaction. 
Through exact diagonalization  (ED) of 
the effective Hamiltonian projected onto the half-filled second moir\'e  band,
we identify an incompressible insulating state with six-fold ground state degeneracy and 1/2  quantized
many-body Chern number, signaling a non-Abelian type FCI.  
The  non-Abelian state is the ground state for the moir\'e MoTe$_2$ system 
for a  range of smaller twist angles, 
with the phase boundary tunable by the strength of the Coulomb interaction.
Furthermore,  we  exclude CDW order by demonstrating a nearly featureless density structure factor for the ground state. These findings imply that the even-denominator 3/2  FCI is the most competitive candidate for twisted bilayer MoTe$_2$ systems at smaller twist angles.  

\section{Results}
\subsection{Band topology} 
The low energy physics of a  hole near the single-layer $K$ valley in the twisted bilayer MoTe$_2$ can be described by
the continuum model  
\begin{equation}
h_{\uparrow\mathbf k}=\begin{bmatrix}
		\frac{(\hat{\mathbf{k}}-K_+)^2}{2m^*}+V_+(\hat{\mathbf{r}}) & \Delta(\hat{\mathbf{r}})\\
		\Delta^+(\hat{\mathbf{r}}) & \frac{(\hat{\mathbf{k}}-K_-)^2}{2m^*}+V_-(\hat{\mathbf{r}})
	\end{bmatrix},\label{eq:ham0}
\end{equation}
where $m^*=0.62m_e$ is the effective hole mass of the monolayer, $K_\pm$ are the $K$ valleys of the top and bottom layers that are twisted by an angle $\theta$, and $V_\pm(\mathbf{r})$ and $\Delta(\mathbf{r})$ are respectively the intralayer and interlayer moir\'{e} potentials~\cite{wu2019topological,reddy2023fractional,Yu2023}. The up arrow in $h_{\uparrow\mathbf k}$ represents the spin-valley locking. The Hamiltonian $h_{\downarrow\mathbf k}$ for a hole near the single-layer $K'$-valley is related to Eq.~(1) by the time reversal symmetry.
Following Refs.~\cite{reddy2023fractional,mao2023lattice,jia2023moire, xwzhang2024,xu2024multiple}, the moir\'e superlattice of the realistic MoTe$_2$ can be simulated by tuning parameters ($V_1, V_2, W_1, W_2, \phi$)  as intra- and inter-layer  first and second harmonics, and the phase
difference $\phi$ between two layers. 
$V_\pm(\mathbf{r})=-2V_1\sum_{j=1,3,5}\cos{(\mathbf{g_j^1}\cdot \mathbf{r}\pm \phi)}-2V_2\sum_{j=1,3,5}\cos{(\mathbf{g_j^2}\cdot \mathbf{r})}$ and
$\Delta(\mathbf{r})=W_1\sum_{j=1,2,3}e^{i\mathbf{q_j^1}\cdot \mathbf{r}}+W_2\sum_{j=1,2,3}e^{i\mathbf{q_j^2}\cdot \mathbf{r}}$,
where  $\mathbf{g_j^1}$ ($\mathbf{q_j^1}$) and $\mathbf{g_j^2}$ ($\mathbf{q_j^2}$) are the momentum differences between the nearest and second-nearest plane wave bases within the same layer
(different layers). 

\begin{figure}
\includegraphics[width=0.98\linewidth,angle=0]{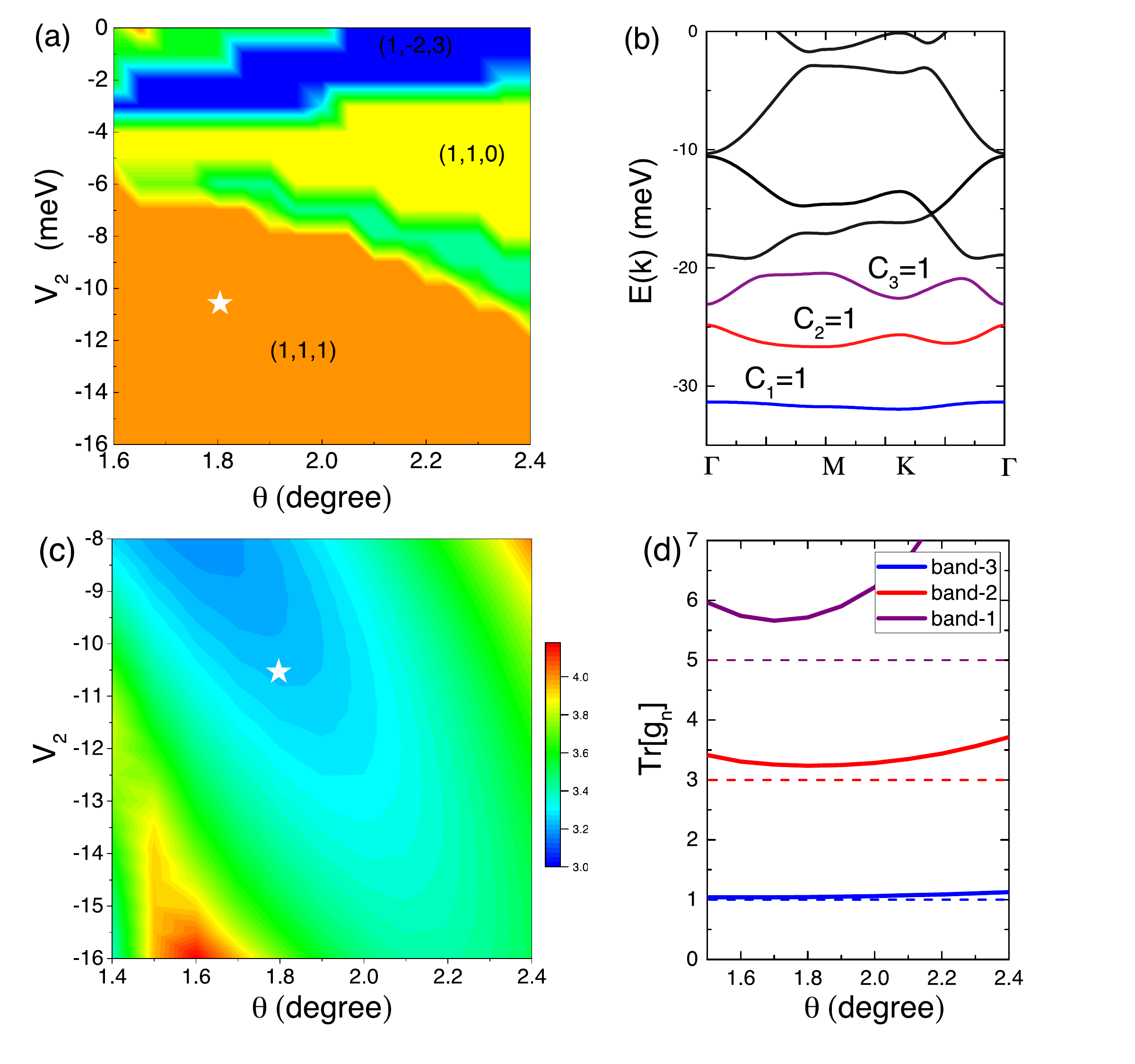}
\caption{ \textbf{Band Topology.} (a)
Band structure diagram of moir\'e MoTe$_2$ as a function of intralayer $V_2$ moir\'e potential strength and the twisted angle $\theta$ for fixed  $V_1=17.5$ meV and $W_1=-6.5$ meV, $W_2=11.75$ meV and $\phi=-58.49^o$.  Different phase regions are distinguished by the Chern numbers of the lowest three moir\'e bands $(C_1,C_2,C_3)$ (as labeled in black color). 
(b) Hole bands for the typical parameters (white star dot in (a)):  $\theta=1.8^o$ and  $V_2=-10.5$ meV,  which will be used for  the many-body simulations. The lowest three moir\'e bands are labeled by their colors, each of which carries Chern number $C=1$ for the $K'-$valley (with opposite Chern numbers in the $K-$valley). 
(c) Heatmap of quantum metric trace condition $\int_\mathrm{B.Z.} \mathrm{tr}[g(\mathbf{k})]/2\pi$ ($\equiv \mathrm{tr}[g_n]$)
of the second lowest band as a function of $V_2, \theta$. 
(d) $\mathrm{tr}[g_n]$ of the lowest three bands vary as a function of twisted angle $\theta$. As a reference, the quantum metric of $n$th Landau level is $2n+1$, denoted as dashed lines in the figure.
}\label{fig:band} 
\end{figure}

The moir\'e potentials from second harmonics are crucial
for accurately capturing higher energy bands from DFT calculations~\cite{jia2023moire}.  
We calculate Bloch wave functions of moir\'e bands and obtain phase diagrams (Fig. \ref{fig:band}(a1,b1))  for  the band Chern numbers in a given valley in terms of the intralayer potential parameter $V_2$ and twisted angle $\theta$. 
The topological character of moir\'e bands depends on $V_2$ and $\theta$ in a complex way and shows several distinct regimes. 
Remarkably, we identify a phase regime where each of the lowest three consecutive moir\'e bands has Chern number $C=1$~\cite{note}. 
For one set of representative parameter values $V_1=17.5$ meV, $W_1=-6.5$ meV, $\phi=-58.49^o$, $V_2=-10.5$ meV, and $W_2=11.75$ meV (white star in Fig. \ref{fig:band}(a)), we show the resulting band structure along high-symmetry lines in the moir\'e Brillouin zone  in Fig.\ref{fig:band}(b). These lowest three topological bands are well separated from other higher energy bands.  
Decreasing the magnitude of $V_2$ leads to  transitions to other band topology regimes. We notice that our parameters are more suitable for small twist angles~\cite{ahn2024landau}. Similar physics is found by slightly changing parameters $\phi$ and $W_2$.

The phase regime with three consecutive Chern bands of $C=1$ provides a suitable regime for targeting
topological physics of excited moir\'e bands. Interestingly, via the calculation of quantum metric \cite{Roy2014}, we identify that the lowest band faithfully resembles the lowest Landau level, and the second lowest band also has a trace condition
close to the first-excited Landau level (see Fig. \ref{fig:band}(c,d)). Specifically, the trace condition  $\int_\mathrm{B.Z.} \mathrm{tr}[g(\mathbf{k})]/2\pi$ for   Fubini-Study (FS) metric $g(\mathbf{k})$ 
of the second lowest band demonstrates a non-monotonic dependence on the twisted angle $\theta$, which shows a minimum around $\theta \sim 1.8^o$ with a value around $3.2$ slightly larger than  the value for the first excited Landau level ($2n+1=3$ for $n=1$) as shown in Fig. \ref{fig:band} (d).   
Next we will explore
around the optimal  parameters
marked as the white stars in Fig. \ref{fig:band}(a,c)) for numerical characterizations of the half filling $\nu=1/2$ many-body state of the second moir\'e band. We believe the physics shown below should be generic and representative for the experimental systems.


\subsection{Many-body Interaction}
The two-body interaction takes the form  
\begin{align}
H_2=\sum_{n_{1,2},m_{1,2}}&\sum_{\sigma_1,\sigma_2}\sum_{\mathbf{k}_{1,2},\mathbf{q}}
V_{n_1n_2m_1m_2\sigma_1\sigma_2}(\mathbf k_1,\mathbf k_2,\mathbf{q}) \times \nonumber \\
&\hat{c}^\dagger_{n_1\sigma_1\mathbf k_1} \hat{c}^\dagger_{n_2\sigma_2\mathbf k_2} \hat{c}_{m_1\sigma_2[\mathbf k_2-\mathbf q]} \hat{c}_{m_2\sigma_1[\mathbf k_1+\mathbf q]},
 \end{align}
where the summation is over all band index $n_{1,2}$, spin/valley index $\sigma_{1,2}$ and momenta $\mathbf{k}_{1,2}$ and $\mathbf{q}$.   $\hat{c}^\dagger_{n\sigma\mathbf k}$ creates a hole in the Bloch eigenstate of $h_{\sigma k}$ with  subband index $n$, spin/valley $\sigma$ and momentum $\mathbf k$ in the moir\'e Brillouin zone. $[{\mathbf k}]={\mathbf k}-{\mathbf g_k}$ where the operator $ [ $  $ ]$ takes $\mathbf {k}$ to its reduced vector in the moir\'e Brillouin zone and $\mathbf {g_k}$ is a moir\'e reciprocal lattice vector. The Bloch state interaction matrix elements are
\begin{align} 
V_{n_1n_2m_1m_2\sigma_1\sigma_2}(\mathbf k_1,\mathbf k_2,\mathbf q)=&\frac{1}{2A}V(\mathbf{q}) f^{n_1m_2\sigma_1}(\mathbf k_1, \mathbf q) \times \nonumber  \\
& f^{n_2m_1\sigma_2}(\mathbf k_2, -\mathbf q),
\end{align}
where $A$ is the area of the system, $\mathbf q$ is the momentum transfer of the scattering process, and the  Coulomb interaction is set to be $V(\mathbf{q})=\frac{ e^2}{\varepsilon_0\varepsilon q} (1-\delta_{q,0})$ with the elementary charge $e$, vacuum permittivity $\varepsilon_0$ and dielectric constant $\varepsilon$. The form factor between $n$ and $m$ subbands is defined as
\begin{align}
f^{nm\sigma}(\mathbf k,\mathbf q) = \langle n\sigma\mathbf k| e^{-i \mathbf q \cdot \mathbf r}|m \sigma [\mathbf k+ \mathbf q] \rangle,
\end{align}
where $|n \sigma \mathbf k \rangle$ is the Bloch eigenstate of $h_{\sigma \mathbf k}$ in Eq. \ref{eq:ham0}.

\begin{figure}
   \includegraphics[width=0.45\textwidth,angle=0]{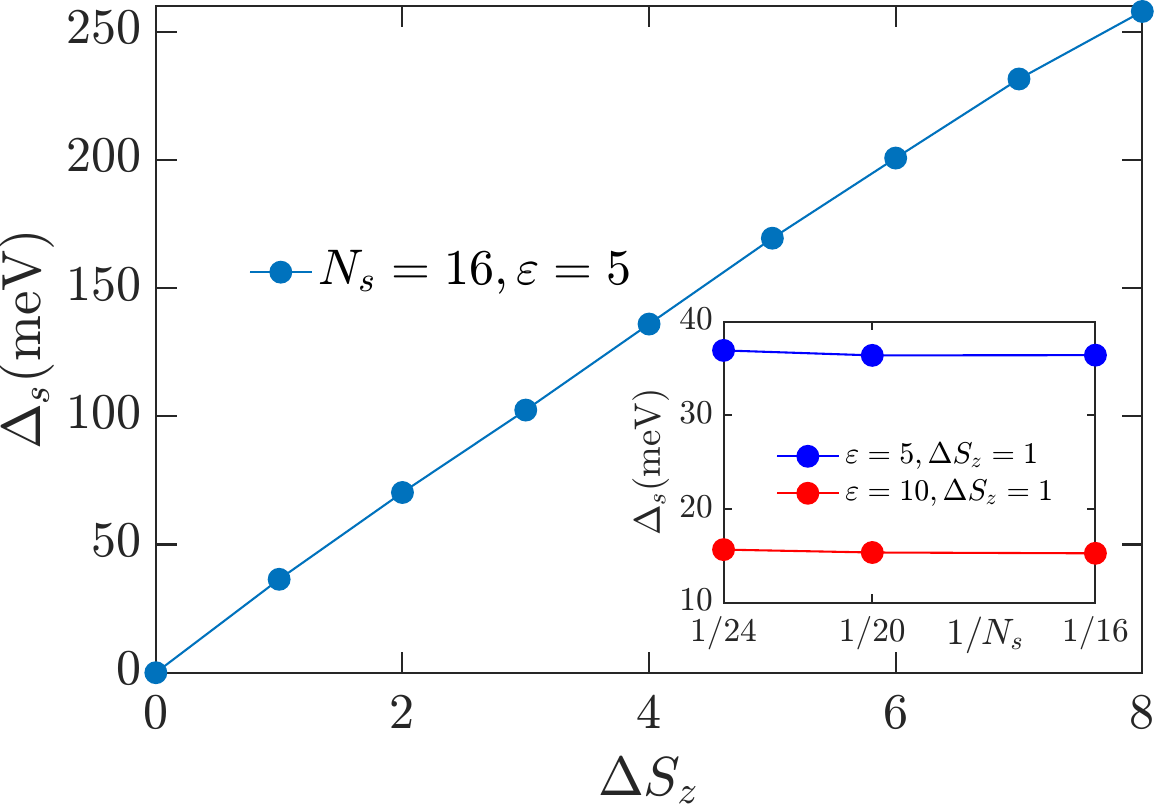}
   \caption{ \textbf{Spin excitation gap.}    The spin gap $\Delta_s=E_0(S_z)-E_0(S_\text{zmax})$ for an $N_s=16$ system 
   at total filling $\nu=3/2$ and $\varepsilon=5$, as a function of the deviation of the total spin from the full polarization $\Delta S_z=S_\text{zmax}-S_z$, where $S_\text{zmax}=N_h/2$ and $E_0(S_z)$ is the lowest energy in the spin $S_z$ sector. The inset shows spin gap $\Delta_s$ at $\varepsilon=5, 10$ for $\Delta S_z=1$ and different system sizes. These indicate a ferromagnetic ground state in the thermodynamic limit.
   The band parameter set is defined in the caption of Fig. \ref{fig:band}. 
   } \label{fig:spingap}
\end{figure}

\subsection{Spontaneous spin polarization at $\nu=3/2$}
Since the numbers   of spin-$\uparrow$ and $\downarrow$ holes $N_\uparrow$ and $N_\downarrow$ are separately conserved, we can perform ED calculation for all spin configurations with different total $S_z=(N_\uparrow-N_\downarrow)/2$. 
We carry out ED calculation for interacting systems with a total Hamiltonian $H=\sum_{n\sigma \mathbf k} \epsilon_{n\sigma}(\mathbf k)\hat{c}^\dagger_{n\sigma\mathbf k}\hat{c}_{n\sigma\mathbf k}+H_2$ at $\nu=3/2$
within the Hilbert space of the time-reversed pair ($\sigma=\uparrow,\downarrow$) of the lowest two bands ($n=1,2$). When a band is fully occupied, the interactions between it and other bands are treated based on Hartree-Fock approximation. 
$\epsilon_{n\sigma}(\mathbf k)$ is the Bloch state energy of the $n$-th moir\'e band with spin $\sigma$. To obtain the energy spectrum, we study finite systems with discrete momentum points ${\bf k}=k_x{\bf T}_1+k_y{\bf T}_2$ with ${\bf T}_1, {\bf T}_2$  being unit vectors of crystal momentum, and $k_{x(y)}=1, ..., N_{x(y)}$ for system size $N_s=N_x\times N_y$  and particle number $N_h=N_\uparrow+N_\downarrow=\nu N_s$.

The $S_z$ of the ground state is always found to satisfy $S_z=S_\text{zmax}=N_h/2$ for $N_s=16-24$, representing spontaneous time-reversal symmetry breaking and FM. We further identify spin gaps $\Delta_s= E_{0}(S_z)-E_{0}(S_\text{zmax})$ as the energy difference between the FM ground state and the lowest energy state 
in spin sector $S_z$, which is a function of 
spin-flip $\Delta S_z=S_\text{zmax}-S_z$ as shown in Fig.~\ref{fig:spingap} for the twist angle $\theta=1.8^o$. A finite spin gap is found  for all
spin sectors for $N_s=16$ and $\varepsilon=5$ and it grows with the spin-flip $\Delta S_z$.  Furthermore,  as shown in the inset of Fig.~\ref{fig:spingap}, 
the $\Delta S_z=1$ spin gap is robust with a near constant value  for different  system sizes $N_s=16-24$ and grows with the strength of the Coulomb interaction at $\varepsilon=5$ and 10, indicating a finite spin gap in the thermodynamic limit, consistent with the Ising FM ground state without $SU(2)$ spin-rotational symmetry~\cite{crepel2023anomalous,reddy2023fractional,Yu2023,sheng2024}. In the rest of this work, we will focus on the spin-polarized sector to access much larger systems.

\subsection{Many-body energy spectrum}

Next, we study spin fully polarized systems by considering the following model
\begin{align}
H^{A}=\sum_{\mathbf k} (\epsilon_{2\uparrow} (\mathbf k)+{\Sigma}(\mathbf k))\hat{c}_{2\uparrow\mathbf k}^+\hat{c}_{2\uparrow\mathbf k}+PH_2P, 
\end{align}
where  ${\Sigma}(\mathbf k)$ is the standard Hartree-Fock self-energy due to the interband interaction between holes in the half-filled second moir\'e band and those in the filled lowest moir\'e band (see Supplementary Note 2~\cite{SM}) and the interaction term $H_2$ is projected onto the second spin-up moir\'e band by the projection operator $P$.  

\begin{figure}
   \includegraphics[width=0.46\textwidth,angle=0]{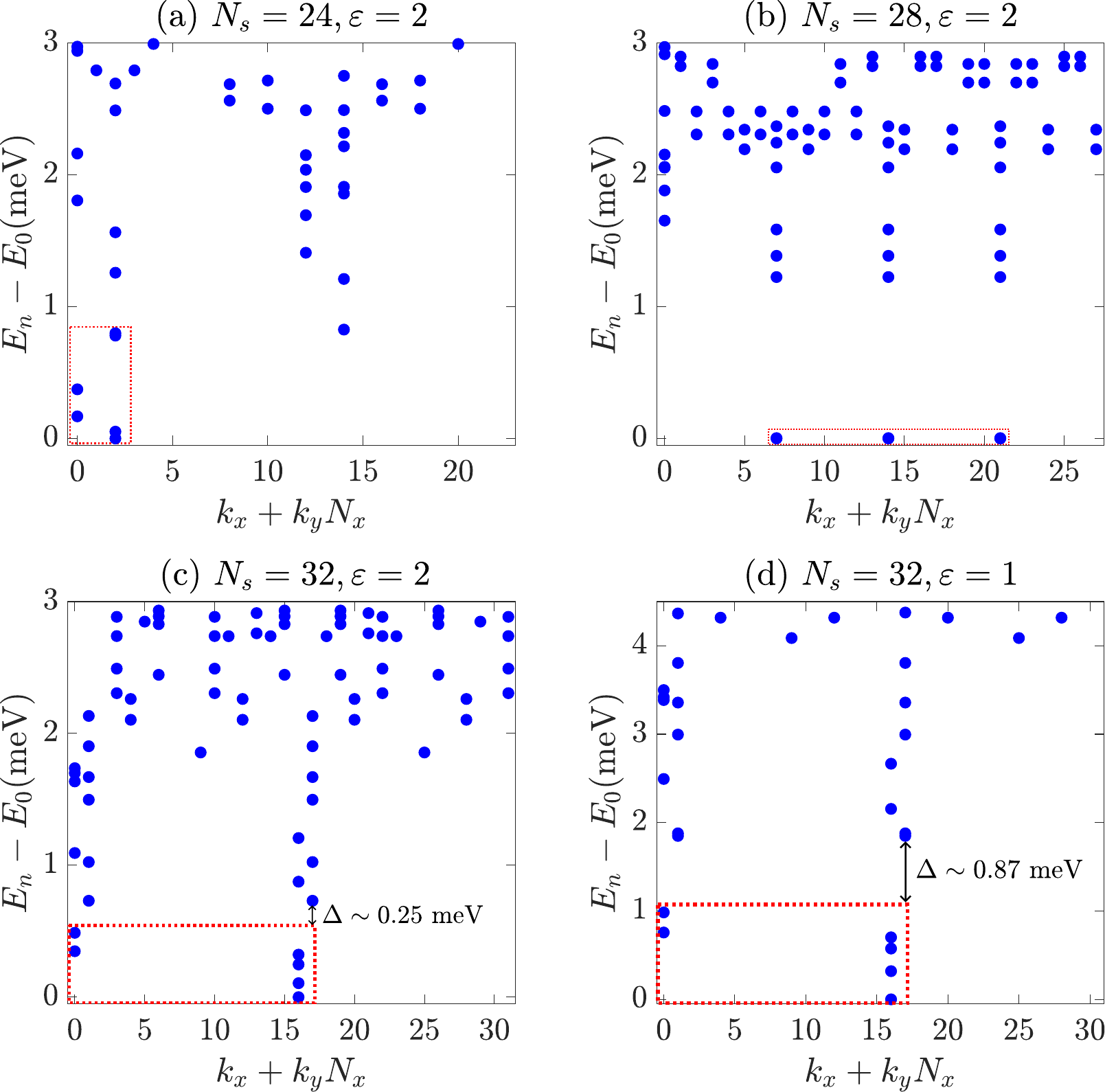}
   \caption{ \textbf{Topological ground state degeneracy.} Energy spectra $E_n$ (shifted by ground state energy) for fractional filling $\nu=1/2$ of the second moir\'e MoTe$_2$ 
    band on various system sizes (a-c) $N_s=24,28,32$ with $\epsilon=2$ and (d) $N_s=32$ with $\epsilon=1$. The ground state manifold matching the degeneracy $N_g=6$ is highlighted by dashed box. 
    The standard Hartee-Fock energy is used  with Hamiltonian $H^A$. 
    The parameter set is defined in the caption of Fig. \ref{fig:band}.
   } \label{fig:EDspectrum}
\end{figure}

One primary feature of a topologically ordered phase is the degenerate ground state manifold protected by a finite energy gap. 
The energy spectrum of $H^A$ at $\nu=1/2$ of the second moir\'e band for three system sizes $N_s= 24$, $28$ and $32$ are 
shown in Fig.~\ref{fig:EDspectrum}. 
We  see that, for each system
size, there is a ground state manifold with six-fold quasi-degenerate states (indicated by the dashed box) 
for our systems with even number of electrons.  In particular, the crystal momenta at which these ground states occur match the momenta of non-Abelian $\nu=1/2$ FCI based on the FQH-FCI folding scheme~\cite{regnault2011fractional,bernevig2012translation}, e.g. for $N_s=24$ two ground states occur at momentum $(k_x,k_y)=(0,0)$ while four others are in the $(2,0)$ sector. However,  for $N_s=24$ and $\varepsilon=2$,  the ground state manifold is not separated from the low energy roton excitations\cite{girvin1986}
centered around the special momentum $\bm k=(N_x/2, N_y/2)$ (or equivalently ($\pi, \pi$) point, indicating strong competition  between FCI and CDW states. For $N_s=28$ cluster with $C_6$ rotational symmetry\cite{reddy2023fractional, sheng2024}, we observe a much enhanced gap between the 6-fold degenerated ground states and the other excited states with reduced finite-size effect. 
Remarkably, for a larger system with $N_s=32$, despite its geometry similar to that of $N_s=24$ which allows  low-energy roton excitations at $(\pi, \pi)$, the excitation gap between the 6th and 7th lowest states 
($\Delta=E_6-E_5$ as we refer to the ground state as $E_0$) is robust and grows more than three times from $\Delta=0.25$ meV to around $0.87$ meV when $\varepsilon=2\rightarrow 1$. This indicates that stronger Coulomb interaction drives the system  into a more robust FCI phase. The observed six-fold ground-state degeneracy precisely matches that of the Pfaffian~\cite{Greiter1991,MOORE1991} or anti-Pfaffian state~\cite{SSLee2007,Levin2007}, consistent with the emergence of a non-Abelian Moore-Read state. 


The ED calculation is inevitably limited by the computational capacity (e.g. The Hilbert subspace of the $N_s=32$ has a dimension of about 19 million which is about the limit of the current ED method). A way to check the finite-size effect is to add a twisted boundary condition 
 as a perturbation to the system, where two boundary phases $\theta_x, \theta_y$ are introduced as generalized boundary conditions in the $\mathbf{L}_1$ and $\mathbf{L}_2$ directions ($\mathbf{L}_{1,2}$ are the lattice vectors of the moir\'e finite size system). Under the twisted boundary conditions, the momentum of single particle shifts to $k_{x(y)}\rightarrow k_{x(y)}+
\frac {\theta_{x(y)}} {2\pi}{\bm {T}_{1(2)}} $, providing an effective way
to scan  all momenta  in the Brillouin zone near continuously and identify topological Chern number which will be presented below.

              \begin{figure}
   \includegraphics[width=0.49\textwidth,angle=0]{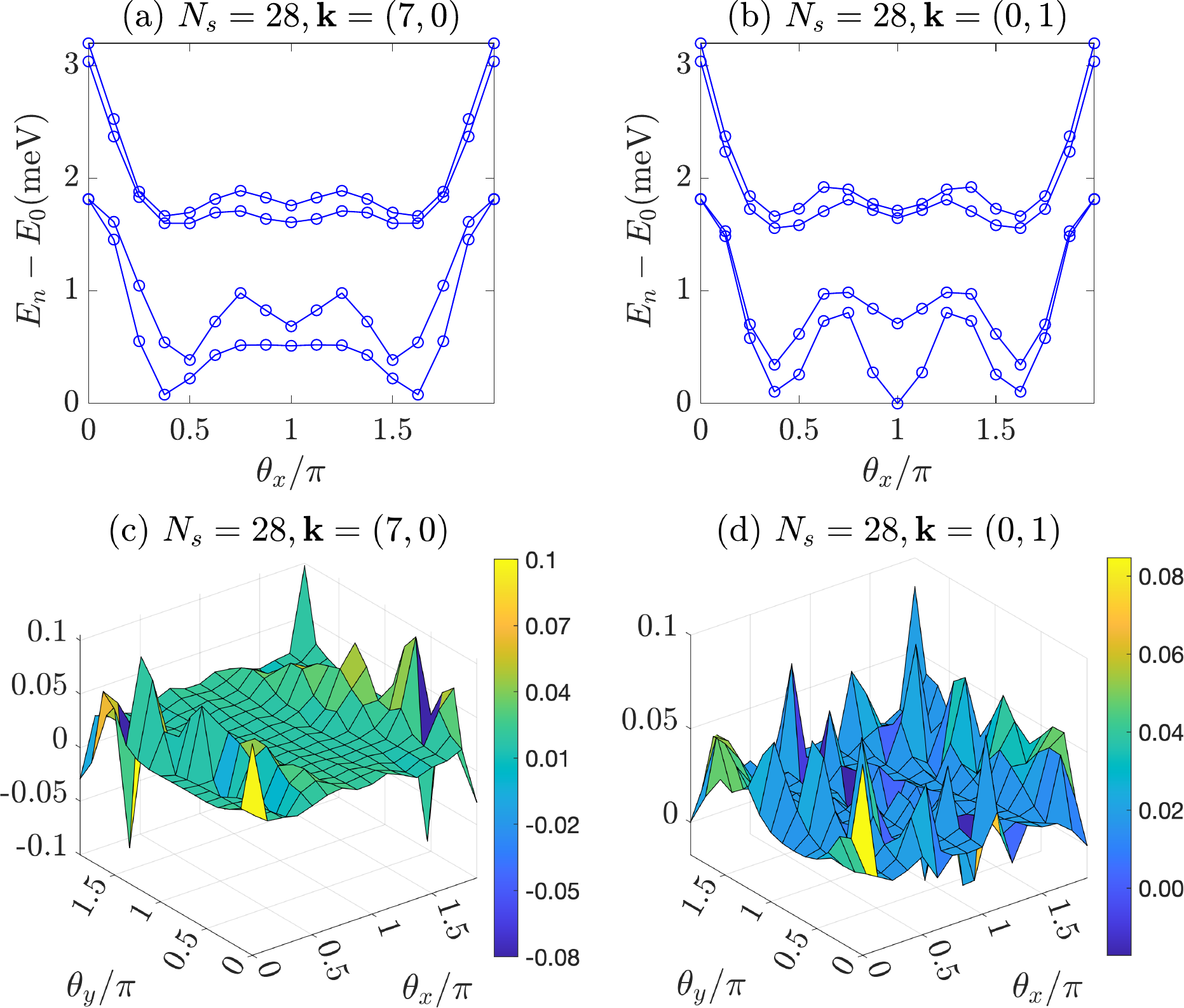}
   \caption{ \textbf{Many-body Chern number.} (a-b) Low-energy spectra $E_n$ (shifted by the ground state energy) at the momentum $\mathbf{k}=(7,0)$ or $(0,1)$ under the variation of the twisted boundary phase $\theta_x$ for $N_s=28$ and 
   $\varepsilon=2$. (c-d) The Berry phase  $F(\theta_x,\theta_y)\Delta\theta_x\Delta\theta_y$ in the discretized  Brillouin zone ($16\times 16$)  for 
   two lowest ground states in each of the two momentum sectors, which sum up to give an exactly quantized total $C=1$.
 } \label{fig:twistedboundary}
\end{figure}

\subsection{Many-body Chern number}
The many-body Chern number (i.e. the Berry phase in unit of $2\pi$) can be defined as an integral invariant using  twisted boundary conditions~\cite{QianNiu1985}: $C={{1}\over{2\pi}}\int d\theta_x d\theta_y
F(\theta_x,\theta_y)$, where the Berry
curvature is given by $F(\theta_x,\theta_y)=\rm{Im}
\left(\left\langle {{\partial
		\Psi}\over{\partial\theta_y}}\Big{|}{{\partial
		\Psi}\over{\partial\theta_x}}\right\rangle -\left\langle {{\partial
		\Psi}\over{\partial\theta_x}}\Big{|}{{\partial
		\Psi}\over{\partial\theta_y}}\right\rangle\right)$. 
  A significant difference between FCI in topological band and FQH state is that, a FCI does not possess the continuous translational symmetry.
  As a consequence,  a FCI phase in the topological band may have
  a quantized topological Chern number different from its filling number~\cite{kol1993fractional, sheng2024}.  Thus the many-body Chern number plays an important role for characterizing the nature of a FCI phase and provides predictions for Hall conductance quantization through the relation $\sigma_H=Ce^2/h$\cite{sheng2003, sheng2011fractional}.

As shown in Fig.~\ref{fig:twistedboundary}(a, b), when tuning the boundary phase $\theta_x$  for a system with  $N_s=28$ and $\varepsilon=2$, the lowest two-fold nearly degenerated ground states in momenta $(7,0)$ and $(0,1)$ sectors are generally separated from the excited states by direct gaps 
,  which allows an accurate determination of the many-body topological Chern number\cite{sheng2003, sheng2011fractional} despite larger energy variances when non-zero boundary phases are present. 
We discretize the boundary phase space into $M_c\times M_c$ square meshes 
 (with $M_c\ge 16$) and numerically obtain the Berry curvature $F(\theta_x,\theta_y)$ for each square~\cite{sheng2011fractional,Fukui2005,wang2011}. 
We show the total Berry phase, which is proportional to Berry curvature for each square mesh in Fig. \ref{fig:twistedboundary}(c,d) at $N_s=28$ and $\varepsilon=2$ with momentum $(7,0)$ and $(0, 1)$, respectively.
By integrating the Berry curvatures over the boundary phase space, we find  total Berry phase of $2\pi$ or quantized total Chern number $C=1$ for the two lowest ground states in both momentum sectors. 
This result confirms that  each nearly degenerate ground state~\cite{sheng2003} carries a fractional Chern number $C=1/2$ and the Hall conductance is quantized at $\sigma_{H}=\frac 1 2 e^2/h$ (the sign of the Chern number is the same as the band Chern number and an additional contribution of
$e^2/h$ from the filled lowest band should be taken into account to compare with experimental observations). 

\begin{figure}[b]
   \includegraphics[width=0.48\textwidth,angle=0]{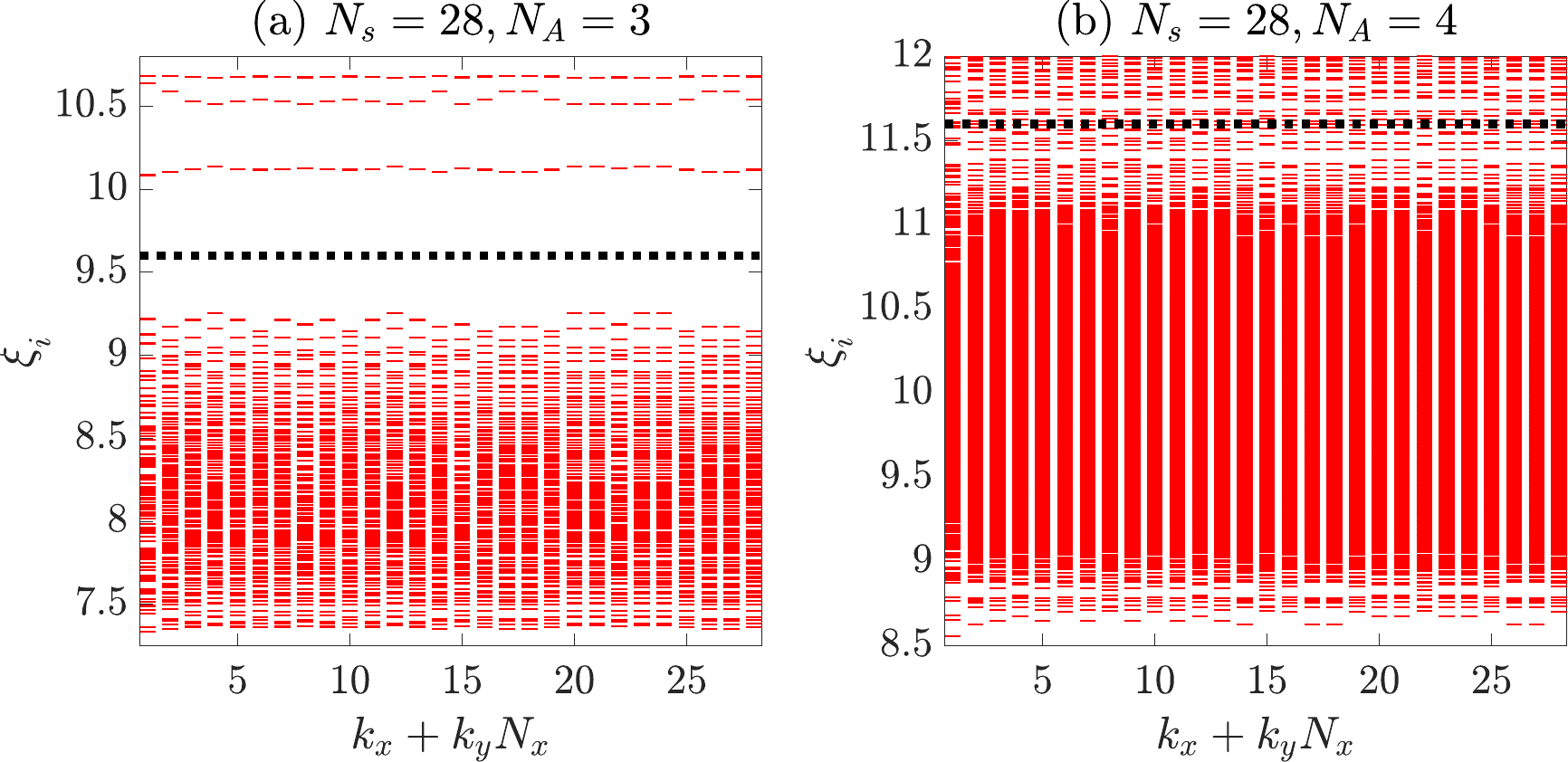}
   \caption{ \textbf {The PES and counting.}  The $N_s=28$ system is  partitioned into  $N_A$ and $N_h-N_A$ particles respectively to obtain PES with (a) $N_A=3$, with 3192 states below the entanglement gap and (b) $N_A=4$, the entanglement gap closes.
   The density matrix is defined as $\rho =\frac{1}{N_g} \sum_{n=1}^{N_g} | \Psi_n\rangle \langle \Psi_n|$, where $|\Psi_n\rangle$ denotes the $N_g=6$ 
   quasi-degenerate ground states.   } 
  \label{fig:PES}
\end{figure}

\subsection{Non-Abelian quasiparticle statistics}

After determining the topological nature of the ground-state manifold, we further inspect the non-Abelian quasiparticle statistics through the entanglement spectroscopy with the particle-cut entanglement spectrum (PES)  \cite{Sterdyniak2011}. By dividing the whole system into $N_A$ and $N_h-N_A$ particles, we identify a clear entanglement gap separating the low-lying PES levels from higher ones (Fig. \ref{fig:PES}(a))
for $N_A=3$ and $N_s=28$. 
There are $3192$  spectrum levels below this gap exactly matching the typical counting of quasiparticle excitations resulting from the generalized Pauli principle of the non-Abelian (anti-)Pfaffian state \cite{regnault2011fractional} (at most 2 particles in 4 consecutive orbitals). 
A similar result is found for $N_s=24$ system with 1952 states below the entanglement gap at $N_A=3$.
Based on the correspondence between the PES and the quasiparticle excitations~\cite{Sterdyniak2011}, this finding further indicates the non-Abelian nature of the half-filled state. 
Furthermore, we find the gap in PES shrinks for the case of $N_A=4$ (Fig. \ref{fig:PES}(b)), consistent with observations in other non-Abelian candidates\cite{reddy2024nonabelian}, which may indicate important contributions from other higher level quasi-particles  or from other low energy competing  states.

\subsection{Structure factors}
\begin{figure}
\includegraphics[width=0.485\textwidth,angle=0]{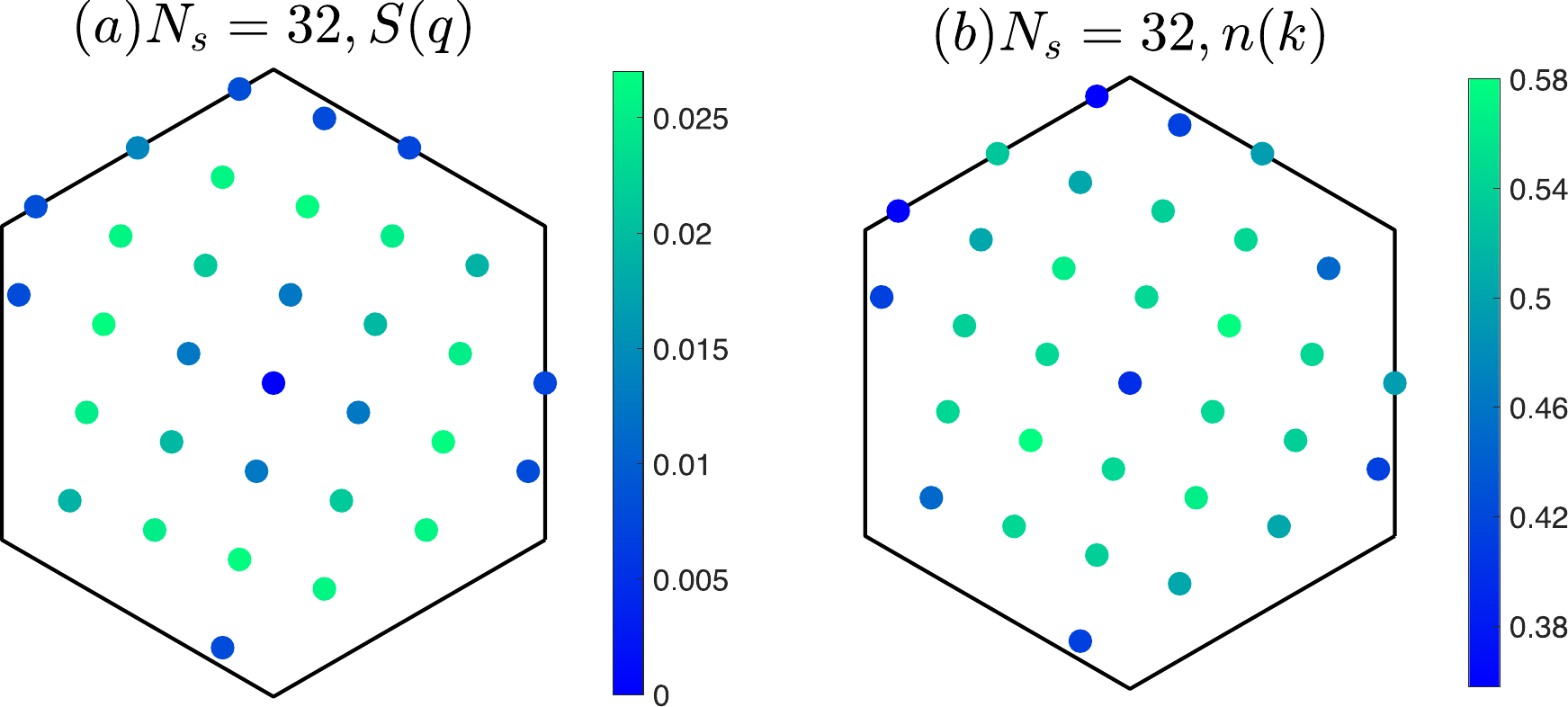}
   \caption{ \textbf{Static Structure Factor.} (a) Density structure factor $S(\mathbf q)$ and (b) momentum-space hole distribution $n(\mathbf k)$. Here we set $\varepsilon=2$ and $N_s=32$. 
   } \label{fig:nk}
\end{figure}

We further investigate the CDW instability. We calculate the projected structure factor as 
\begin{align}
S(\mathbf q) = \frac{\langle \overline{\rho}({\mathbf q}) \overline{\rho}({-\mathbf q}) \rangle }{N_s} 
-\frac {\langle {\overline{\rho}}(0)\rangle^2} {N_s} \delta_{\bm{q},\bm{0}},
\end{align}
where $\overline{\rho}(\mathbf q) = \sum_i e^{-i\mathbf q \cdot \mathbf r_i}$ is the projected density operator  and $\mathbf r_i$ is the hole position.  
In Fig. \ref{fig:nk}(a), $S(\mathbf q)$ exhibits very weak feature with slightly enhanced intensity around the middle 
of  the Brillouin zone, but lacks  strong peaks as expected for a crystal, excluding  the CDW as an instability at half-filling of the second moir\'e band.
In comparison, the momentum space occupation numbers $n(k)$ (Fig. \ref{fig:nk}(b)) are near constant with smaller variance with $\mathbf k$, consistent with a topological  state.

\begin{figure}
   \includegraphics[width=0.485\textwidth,angle=0]{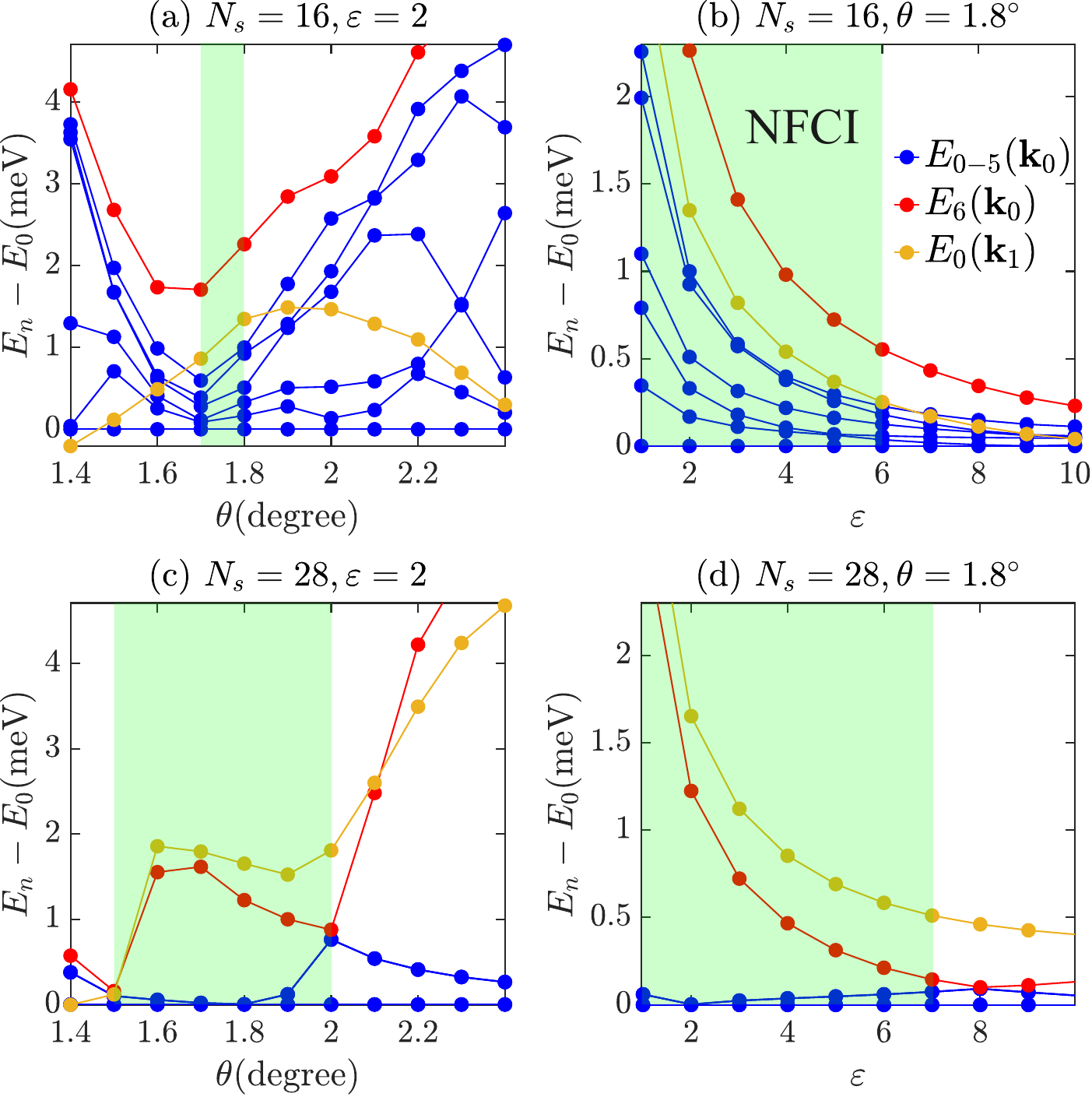}
   \caption{ \textbf{Robustness of non-Abelian FCI.} Energy spectra $E_n$ (shifted by ground state energy) 
   at fractional filling $\nu=1/2$ of the second moir\'e band of MoTe$_2$ at various $\varepsilon$
   and $\theta$. The parameter regimes for the non-Abelian FCI are labeled by the green shades. 
   We show 6-lowest states ($E_{0-5}$) and   excited state ($E_6$) with $\bf {k_0}=(0,0)$ ($N_s=16$) or  M points ($N_s=28$); and lowest  excited state $E_0$ with $\bf {k_1}$  at M ($N_s=16$), or (0,0) point ($N_s=28$). (a,b) Energy spectra and non-Abelian FCI phases for $N_s=16$ by varying $\theta$ or $\varepsilon$, respectively.  The symbols for  ground states and excited states are defined in (b) and applicable to all figures; (c,b) Energy spectra and non-abelian FCI phases for $N_s=28$. 
   } \label{fig:robustness}
\end{figure}

\subsection{Robustness of non-Abelian FCI}
At last, we discuss the robustness of the non-Abelian FCI state. Here we mainly focus on two factors: twisted angle $\theta$ that controls the single-particle band structure and the dielectric constant $\varepsilon$ that controls the Coulomb interaction strength. Fig. \ref{fig:robustness} demonstrates the evolution of many-body energy spectra by varying $\theta$ and $\varepsilon$, respectively. 
By fixing $\varepsilon=2$, we find that the non-Abelian FCI state is favored around the optimal angle $\theta \sim 1.8^o$ at $N_s=16$ indicated by the green shaded area, where the ground state has approximately six fold degeneracy, separated with other excited states by a small gap (Fig. \ref{fig:robustness}(a)).   The  non-Abelian FCI phase expands to cover the regime 
$1.5^o<\theta< 2.0^o$ for a larger system size $N_s=28$ (Fig. \ref{fig:robustness}(c))
, with an excitation gap away from the phase boundaries much larger than the splitting of the six-fold near degenerating ground states.
At fixed twisted angle $\theta=1.8^o$, we find that the non-Abelian  state is established in the strong interaction regime (small dielectric constant $\varepsilon\lesssim 7.0$ see Fig. \ref{fig:robustness}(b,d)). This supports that strong correlation is necessary to stabilize more robust non-Abelian FCI phase. 

\section{Discussion}
We have performed a systematic study on the half-filled second moir\'e band of the
twisted MoTe$_2$ systems. 
We offer numerical evidence of an emergent 
non-Abelian FCI phase with six-fold ground state degeneracy. 
By studying  various system sizes,
we conclude that this ground state manifold is robust and protected by a finite energy gap.
Furthermore, the  folding of momentum quantum numbers of the ground states, and the half integer quantized  topological Chern number demonstrate
this even denominator topological state to be a non-Abelian FCI state.
The featureless density structrue factor supports the absence of a co-existing crystal order at small twist angles $1.5^o<\theta<2.0^o$.  
Such a FCI state is seen to grow more robust with the increase of system size, similar to the
non-Abelian states of the first excited Landau level, which makes the experimental detection promising. Additional results with modified  kinetic energies connecting the realistic model we study
with flatband limit 
demonstrate similar physics, which are presented in  Supplementary Note 1~\cite{SM}).

Although the many-body
spectrum and  Chern number quantization of the even denominator 
FCI of twisted MoTe$_2$ system fit the expectation of Moore-Read Pfaffian (anti-Pfaffian) state, which type of topological order is realized for  this novel state may require future study. Our results suggest both Pfaffian and anti-Pfaffian as the most likely candidates.
Historically, the nature of the half-filled first excited Landau level quantum Hall effect is a long-standing issue, i.e. several candidates such as the Pfaffian state, anti-Pfaffian state and particle-hole symmetric Pfaffian state, are competing states if particle-hole symmetry breaking effects are neglected.  In the twisted TMD systems, the particle-hole symmetry is  explicitly broken by the non-zero kinetic dispersion and orbital dependent Berry curvature, which makes the particle-hole Pfaffian less likely to be a competing phase in our system. 
However, as the quantum metric of the second moir\'e band is very close to the first excited Landau level~\cite{reddy2024nonabelian,ahn2024landau,liu2024theory}, this many-body state may be primarily inherited from the particle-particle interaction rather than the kinetic energy. With the emergence of the many-body translational symmetry\cite{bernevig2012translation} for this FCI, there is a possible emergence of particle-hole symmetry.
We believe these open questions can be addressed in the future which will  advance our theoretical understanding of non-Abelian physics.
Our work  makes a concrete prediction for realizing such a non-Abelian phase in  twisted moir\'e systems at half filling of the second moir\'e band, which will stimulate experimental discoveries along this direction.
Interestingly, a recent work considering spinful systems at total filling $\nu=3$  suggests that a spin unpolarized state is possible by weakening the short-range interaction between holes with opposite spins\cite{abouelkomsan2024}. This leaves the opportunity for time-reversal invariant quantum spin Hall phases to emerge\cite{kang_observation_2024} or to compete with Ising spin polarized states, which we leave  for  future studies.



\section*{Methods}
The topological band model for twisted bilayer MoTe$_2$ system is numerically simulated using the exact diagonalization (ED). In this paper, the system size varies from $N_s=16$ to $32$. For each system size, the calculation can be paralleled by using the total momentum of particles $(k_{x},k_y)$. 
The Hilbert subspace of the $N_s = 32$ has the dimensions of about 19 million which is about the limit of the current ED method.

\subsection{Calculation of Chern number for many-body state}
The Chern number of a many-body state can be obtained as an integral invariant using  twisted boundary conditions~\cite{QianNiu1985}: $C={{1}\over{2\pi}}\int d\theta_x d\theta_y
F(\theta_x,\theta_y)$, where the Berry
curvature is given by $F(\theta_x,\theta_y)=\rm{Im}
\left(\left\langle {{\partial
		\Psi}\over{\partial\theta_y}}\Big{|}{{\partial
		\Psi}\over{\partial\theta_x}}\right\rangle -\left\langle {{\partial
		\Psi}\over{\partial\theta_x}}\Big{|}{{\partial
		\Psi}\over{\partial\theta_y}}\right\rangle\right)$.
To determine the Chern number accurately, we divide the boundary-phase unit cell into $M_c\times M_c$  meshes with $M_c\geq 16$. The Berry curvature is then given by the Berry phase of each mesh divided by the area of the mesh. The Chern number is obtained by summing up the Berry phases of all the meshes.

\subsection{Projected structure factor}
The  projected structure factor is defined as $S(\mathbf q) = \frac{\langle \overline{\rho}({\mathbf q}) \overline{\rho}({-\mathbf q}) \rangle }{N_s} 
-\frac {\langle {\overline{\rho}}(0)\rangle^2} {N_s} \delta_{\bm{q},\bm{0}}$,
with $\overline{\rho}(\mathbf q) = \sum_i e^{-i\mathbf q \cdot \mathbf r_i}$ as the projected density operator  and $\mathbf r_i$ is the hole position.  Through analyzing the $\mathbf {q}$
dependence of $S(\mathbf q)$, one can identify or exclude possible CDW order of the ground states.

\section*{Supplementary Information}
The Supplementary information contains more details to support the discussion in the main text.\\
Supplementary Figure 1: Many-body energy spectra for a model with modified kinetic energy connecting realistic twisted MoTe$_2$ system to its flatband limit. \\ 
Supplementary Figure 2: Phase diagram for non-Abelian state with varying the modified kinetic energy.\\ 
Supplementary Figure 3: Renormalized single-particle band structure.\\
Supplementary Figure 4: Many-body energy spectra for a model with modified Hartree-Fock self-energy for even number of particles.\\
Supplementary Figure 5: Many-body energy spectra for systems with odd number of particles.  \\
Supplementary Figure 6: Berry curvature $F(\theta_x,\theta_y)\Delta\theta_x\Delta\theta_y$  for $N_s=28$ system. \\
Supplementary Figure 7: Density structure factor and momentum distribution function for  $N_s=28$ system. \\
Supplementary Figure 8: Entanglement spectroscopy with the particle-cut entanglement spectra  for $N_s=28$. \\
Supplementary Figure 9: Many-body energy spectra at $N_s=16, 24, 28$ and 32 for $\theta=2.1^o$.

\section{Author contributions}
W.Z.\ and D.N.S.\ initiated the project.  
All authors contributed  to the simulations, analysis of the data and writing of the manuscript.

\section{Data availability}
The data that support the plots within this paper and other findings of this study are available in the GitHub repository (\href{https://github.com/cfengno1/Non-Abelian-FQHE-in-Moire-Systems/tree/main/Data}{https://github.com/cfengno1/Non-Abelian-FQHE-in-Moire-Systems/tree/main/Data}).

\section{Code availability}
Source codes for the numerical simulations are available in the GitHub repository (\href{https://github.com/cfengno1/Non-Abelian-FQHE-in-Moire-Systems/tree/main/Code}{https://github.com/cfengno1/Non-Abelian-FQHE-in-Moire-Systems/tree/main/Code}).

\section{Acknowledgement} We acknowledge stimulating discussions with Zhao Liu,  Tiansheng Zeng, Yang Zhang and
Cheng Xu.  D.N.S thanks Liang Fu, A. P. Reddy, A. Abouelkomsan and 
E. J. Bergholtz for collaborations on  Hall-crystals of moir\'e system and helpful discussions.
This work was primarily supported by the U.S. Department of 
Energy, Office of Basic Energy Sciences under Grant No. DE-FG02-06ER46305 (F.C., D.N.S.) for studying  interacting topological systems.  W.Z. is supported by the foundation of Westlake University. W.W.L. is supported by Zhejiang Provincial Natural Science Foundation of China.

\section{ Note added}
We recently became aware of related works on this topic~\cite{wang2024higher,ahn2024landau,xu2024multiple,reddy2024nonabelian}.
While our scope is different, their  results 
on band structures  or many-body energy spectra of related models are qualitatively consistent with our results. 
At the final stage of this work, we learned of a parallel work \cite{Emil2024} that addresses similar issues in a multi-layer graphene-based moir\'e system.

\bibliography{ref}

\begin{thebibliography}{56}
\bibitem [{not()}]{note}%
  \BibitemOpen
  \href@noop {} {\bibinfo {title} {For simplicity, we neglect an overall negative sign of chern numbers for all bands due to spontaneous time-reversal symmetry breaking.}}\BibitemShut {Stop}%
\end{thebibliography}

\clearpage
                                                                  \clearpage
\appendix
\widetext
\begin{center}
	\textbf{\large Supplementary Information for: “Robust non-Abelian  even-denominator fractional Chern insulator in twisted bilayer MoTe$_2$”}
\end{center}
\vspace{1mm}
\maketitle

\renewcommand{\figurename}{Supplementary Figure}
\setcounter{figure}{0} 
\setcounter{equation}{0}

In the Supplementary Information, we provide additional numerical results to support the conclusion we have discussed in the main text.  Note 1 provides understanding of how the realistic twisted MoTe$_2$ system and its flatband limit are connected by gradually reducing the  strength of the kinetic energy term in the continue model.
 Note 2 provides details of renormalized band structure based on interband interaction and introduces model B ($H^B$), 
  which has reduced Hartree-Fock self-energy. Following that, we provide more results
for model B ($H^B$) in Note 3-7. 
Note 3 shows energy spectra for different system sizes  to demonstrate the even-odd effect consistent with non-Abelian (anti-)Pfaffian state.   Note 4 shows the
Berry phases for  systems $N_s=28$ and robust Chern number quantization for the  non-Abelian state.   Note 5 has density
structure factor and momentum distribution function  for  $N_s=28$
system to demonstrate very similar results as $N_s=32$ shown in the main text.
 In Note 6, we provide results on particle partition entanglement spectrum.
These results further support our main conclusion that the 3/2 non-Abelian even denominator fractional Chern insulator is robust.
While all these results  are presented for twist angle $\theta=1.8^{o}$, in Note 7 we show evidence of the non-Abelian  fractional Chern insulator (NFCI)
for twist angle $\theta=2.1^{o}$. 

\begin{figure}
\includegraphics[width=0.6\linewidth,angle=0]{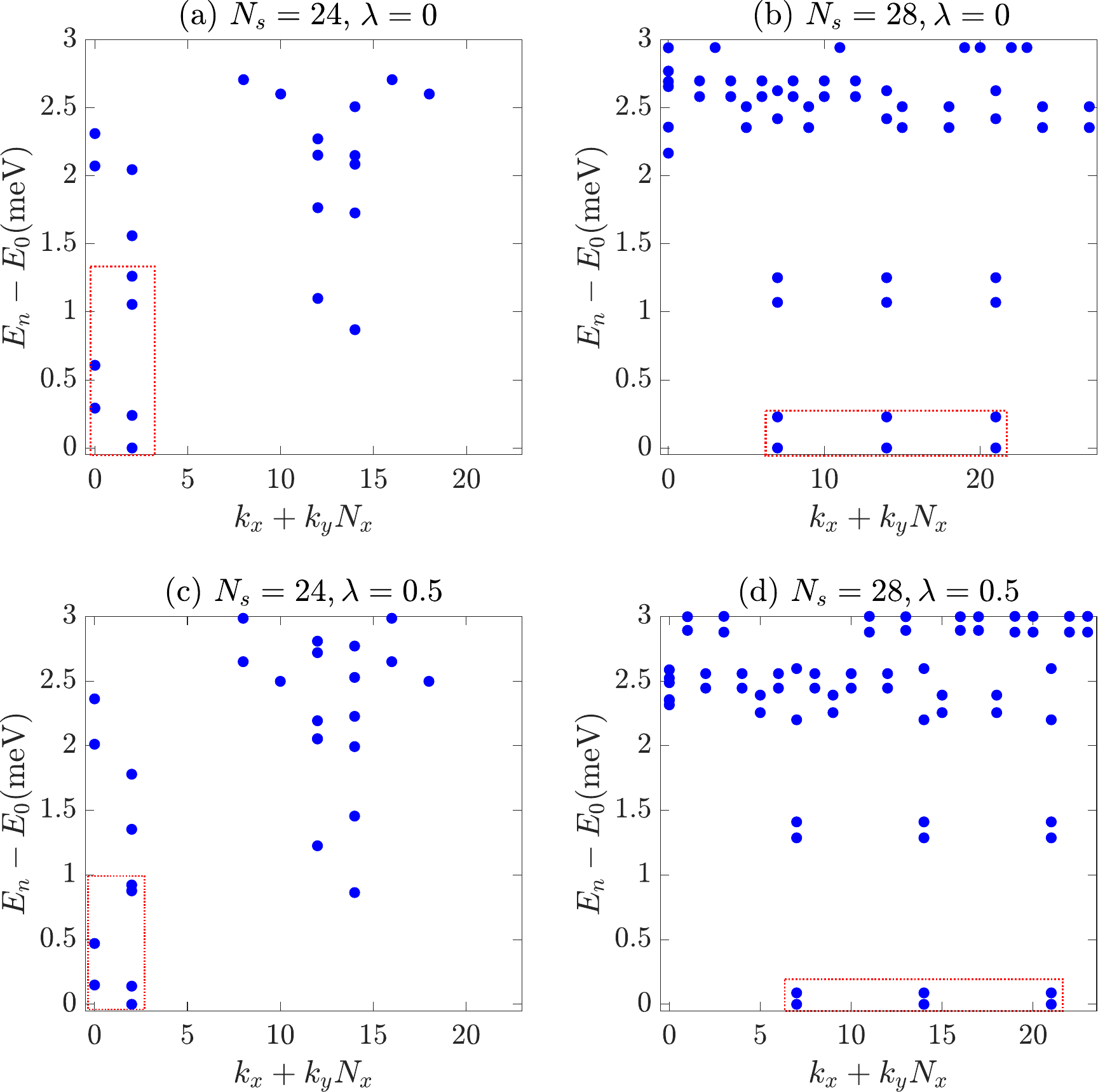}
\caption{ Energy spectra at $N_s=24, 28$ and  $\theta=1.8^o$ for the model in the flat-band limit (a,b) with $\lambda=0$ or with reduced strength of the  kinetic energy (c,d) at $\lambda=0.5$. Other parameters are noted in the caption of Fig.~1 in the main text.
}\label{fig:flat-band} 
\end{figure}

\begin{figure}
\includegraphics[width=0.95\linewidth,angle=0]{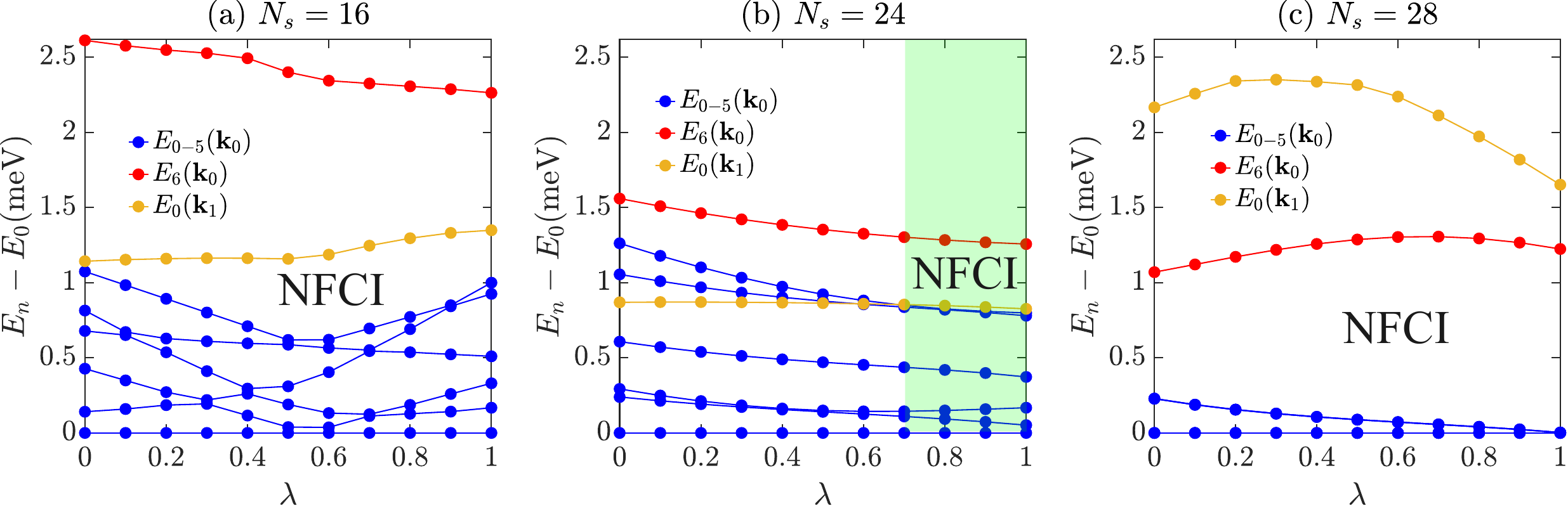}
\caption{
 Energy spectra and NFCI phase for (a)$N_s=16$; (b)$N_s=24$; (c) $N_s=28$.  We show 6-lowest states ($E_{0-5}$) and   excited state ($E_6$) with $\bf {k_0}=(0,0)$ for $N_s=16$, $(0,0)$ and $(N_x/2,0)$ for $N_s=24$,  and three  M points for $N_s=28$; and lowest  excited state $E_0$ from another  momentum sector $\bf {k_1}$ at M point for $N_s=16$, $(N_x/2,N_y/2)$
 for $N_s=24$, and $(0,0)$ for $N_s=28$, respectively. }\label{fig:gap-flat-band} 
\end{figure}

\section{\label{sec:flatband} Supplementary Note 1: Non-Abelian topological phase with reduced kinetic energies: from  flatband limit to realistic moir\'e model for  MoTe$_2$ system}

In the main text, we have studied the model Hamiltonian
\begin{align}
H^{A}=\sum_{\mathbf k} (\varepsilon_0 (\mathbf k)+{\Sigma}(\mathbf k))\hat{c}_{2\mathbf k}^+\hat{c}_{2\mathbf k}+PH_2P, 
\end{align}
where  $\varepsilon_0(\mathbf k) $
is the orbital energy for Bloch states and  ${\Sigma}(\mathbf k)$ is the standard Hartree-Fock self-energy due to interband interaction (see next section) and the interaction term $H_2$ is being projected into second moir\'e band (for simplicity, we do not show explicitly the spin dependence of the band). 
Here we modify the kinetic energy  to $H_k(\lambda)=\sum_{\mathbf k} \lambda(\varepsilon_0 (\mathbf k)+{\Sigma}(\mathbf k))\hat{c}_{2\mathbf k}^+\hat{c}_{2\mathbf k}$, taking $\lambda=0\rightarrow 1$ to connect the flatband limit ($\lambda=0$) to realistic MoTe$_2$ model at $\lambda=1$.

In Supplementary Fig. \ref{fig:flat-band} we show the energy spectrum for $\varepsilon=2$, $N_s=24$ and $28$ at $\lambda=0 $ and $0.5$, respectively.  For $N_s=24$ (Supplementary Fig. \ref{fig:flat-band}(a,c)), we find  six near degenerating ground states from ${\mathbf k}=(0,0), (2,0)$
sectors, which are not separated from roton excitations at other M-points  $(0, N_y/2)$ and $(N_x/2, N_y/2)$.
These energy spectra are similar to the one for the realistic model $H^A$ (or $\lambda=1$) shown in the main text, but with enhanced competition between non-Abelian state and competing charge density  wave (CDW).  This may be understood as following. At the flatband limit ($\lambda=0$), the effective model  with interaction has approximate particle-hole symmetry, which allows a 12-fold near degeneracy for possible Pfaffian or anti-Pfaffian candidate states, and their superpositions. This enlarged degeneracy also makes it more flexible towards developing CDW order to co-eixst with topological order for a finite system, making it harder to determine the nature of the quantum state at thermodynamic limit.  In fact, this particle-hole symmetric flatband limit is similar to the first excited Landau level physics. 
For $N_s=28$, due to larger system sizes and high symmetry of the cluster ($C_6$ rotational symmetry)~\cite{sheng2024},  we find an enhanced six fold ground state degeneracy, which are well separated from all other excitations, in consistent with a non-Abelian FCI state. 
Even in this case, we find possible 12-fold degeneracy for both $\lambda=0.0, 0.5$ below other excited states (in the same momentum sectors as the lowest 6-states), indicating the possible competing physics  at $\lambda=0$ or 
 $\lambda=0.5$.

We further explore the evolution of the energy spectra for $\nu=1/2$ filled second moir\'e band as a function of $\lambda=0\rightarrow 1.0$ for $\theta=1.8^o$ and $N_s=16, 24, 28$ as shown in Supplementary Fig. \ref{fig:gap-flat-band}(a,b,c).
For smaller cluster $N_s=16$ with rotational symmetry,
we find well defined six fold ground state degeneracy in the
$\mathbf k=(0,0)$ sector, in consistent with a non-Abelian FCI phase
for the whole range of the $\lambda$.
However, the gap between a roton excitation ($E_0(\mathbf k_1)$) and six-fold near degenerating states is relatively small, which grows bigger at $\lambda\geq0.5$ side and remains robust
for the realistic model of tMoTe$_2$ at $\lambda=1.0$.
For $N_s=24$, we see much stronger competition between non-Abelian state and CDW order, which may co-exist for this cluster, and we label shaded regime as NFCI for $\lambda\geq0.7$, where 6-fold degenerating non-Abelian states are below other excited states although with very tiny gap.
Importantly, for larger and higher symmetry cluster with $N_s=28$, we find the whole regime as a robust NFCI phase, with excitation gap much larger than the splitting of the six near degenerating ground states.
Thus we conclude, from the flatband limit to realistic tMoTe$_2$ model, the non-Abelian state is the most robust candidate state at $\nu=1/2$ of the second moir\'e band.

\section{Supplementary Note 2: Renormalized band structure and interband interaction}
Here we consider interaction between the first moire band and the second moire band at total filling 
$\nu=3/2$ with polarized spins, by assuming the first band is fully occupied. The interband interaction is written as
       \begin{equation}\label{}   
    \begin{aligned}
       \hat{H}^{\text{inter}}
    =&     \sum_{\substack{ \bm{k}_{1-4}  , n,m         } }  
       \hat{c}^{\dagger}_{n \bm{k}_1 }
       \hat{c}^{\dagger}_{m \bm{k}_2 } 
       \hat{c}_{m \bm{k}_3} 
       \hat{c}_{n \bm{k}_4} 
      V_{  n m m n} (\bm{k}_1, \bm{k}_2, \bm{k}_3, \bm{k}_4 )   \\
      &+    \sum_{\substack{ \bm{k}_{1-4}  , n,m         } }  
       \hat{c}^{\dagger}_{m \bm{k}_1 }
       \hat{c}^{\dagger}_{n \bm{k}_2 } 
       \hat{c}_{m \bm{k}_3} 
       \hat{c}_{n \bm{k}_4} 
      V_{  m n m n} (\bm{k}_1, \bm{k}_2, \bm{k}_3, \bm{k}_4 ) 
    \end{aligned}
    \end{equation}
    where $\hat{c}^+_{m \bf {k}}$ is the  creation operator of a hole in the
    $m$ band with momentum $\bm{k}$ in the moir\'e Brillouin zone. $n (m) =1, 2$ is the band index and $n,m$ are not equal with each other.  
    The matrix element $V_{n_1,n_2,n_3,n_3} (\bm{k}_1, \bm{k}_2, \bm{k}_3, \bm{k}_4 ) $  denotes
    \begin{equation}\label{}  
    \begin{aligned}
            V_{  n_1 n_2 n_3 n_4} (\bm{k}_1, \bm{k}_2, \bm{k}_3, \bm{k}_4 )  &= \frac 1 {2A}\sum_{\substack{ \bm{q}   } }  
       \delta_{ \bm{k}_4 , [\bm{k}_1+ \bm{q}]}   
      \delta_{\bm{k}_3 , [\bm{k}_2 - \bm{q}]}   
      V(\bm{q})f^{n_1,n_4}{(\bm{k}_1,  \bm{q})}
         f^{n_2,n_3}{(\bm{k}_2,  -\bm{q})} 
        \\
    \end{aligned}
    \end{equation}
$\bm{q}$ is the momentum transfer for Coulomb interaction and the operator $ [ $  $ ]$ takes $\bm{k}\pm \bm{q}$ to its reduced vector in the moir\'e Brillouin zone.
  Alternatively 
     \begin{equation}\label{}  
    \begin{aligned}
            V_{  n_1 n_2 n_3 n_4} (\bm{k}_1, \bm{k}_2, \bm{q} )  &=  
       \frac 1 {2A}V(\bm{q})  
           f^{n_1,n_4}{(\bm{k}_1,  \bm{q})}
         f^{n_2,n_3}{(\bm{k}_2, -\bm{q})} 
        \\
    \end{aligned}
    \end{equation}
where $A$ is the area of the system, $V(q)$ is Coulomb interaction in momentum space.
\begin{equation}\label{}  
\begin{aligned}
f^{n_1n_2}(\mathbf k,\mathbf q) = \langle n_1 \mathbf k| e^{-i \mathbf q \cdot \mathbf r}|n_2 [\mathbf k+ \mathbf q] \rangle 
\end{aligned}
    \end{equation}
    where $|n \mathbf k \rangle$ is the Bloch eigenstate of moir\'e system.   

\begin{figure}[H]
\includegraphics[width=0.98\linewidth,angle=0]{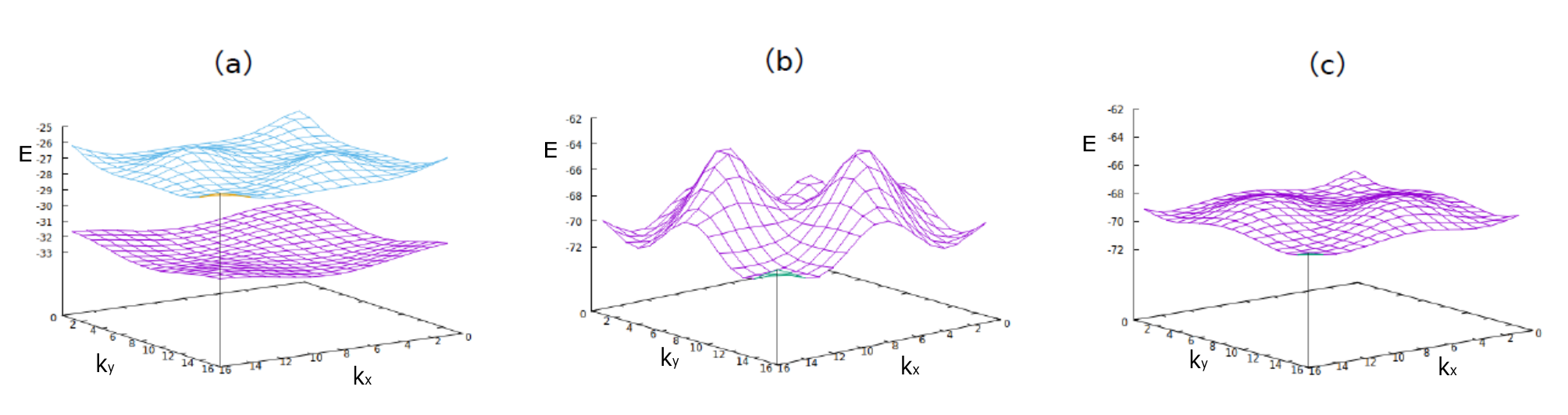}
\caption{(a) Single-particle band structure of the first (purple) and second (blue) moire band. The energy band gap is round $6$meV. (b) The renormalized second moire band by considering the standard Hartree-Fock correction. The renormalied band becomes more dispersive due to the
HF self-energy, with the band fluctuation around $~9$meV.
(c)  The renormalized second moire band by considering the empirical Hartree-Fock self-energy correction. 
The parameters are set to be $V_1=17.5$ meV, $W_1=-6.5$ meV, $\phi=-56.49^o$, $V_2=-11$ meV, $W_2=12$ meV, $\varepsilon=5$. $x,y$-axis are momentum $k_x,k_y$ in the Brillouin zone and $z$-axis is the band energy.
}\label{fig:HF} 
\end{figure}

\begin{figure}[H]
   \includegraphics[width=0.85\textwidth,angle=0]{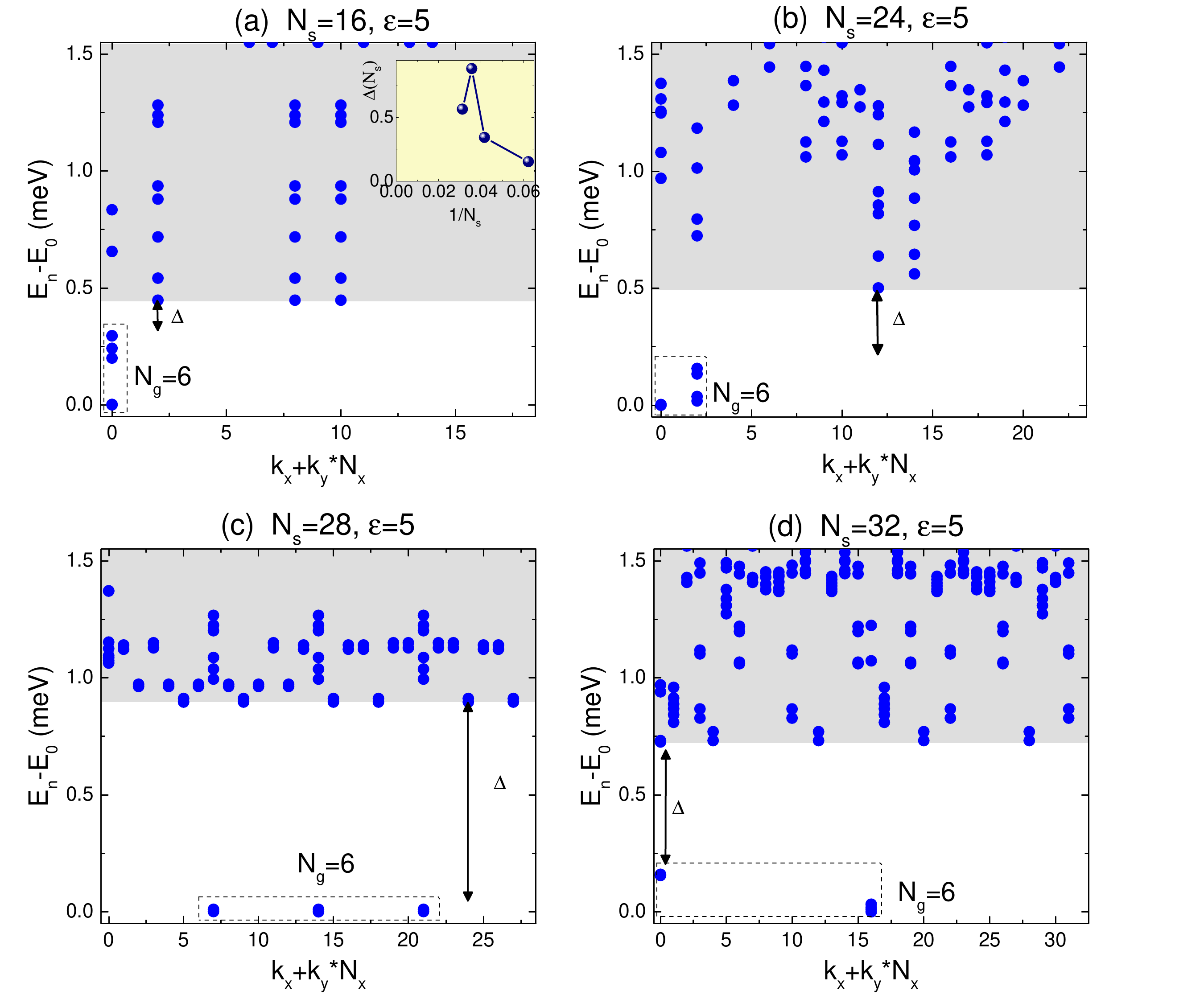}
   \caption{ Energy spectra $E_n$ (shifted by ground state energy) for fractional filling $\nu=1/2$ of the second moir\'e MoTe$_2$ band on  various system sizes (a) $N_s=16$, (b) $24$, (c) $28$, (d) $32$
   by setting $\varepsilon=5$. The ground state manifold matching the degeneracy $N_g=6$ is highlighted by dashed box. 
   The inset shows the energy gap (difference between the  seventh and sixth lowest states) versus $1/N_s$. This is for Model B.
   } \label{fig:spectrum}
\end{figure}

\begin{figure}[H]
\includegraphics[width=0.98\linewidth,angle=0]{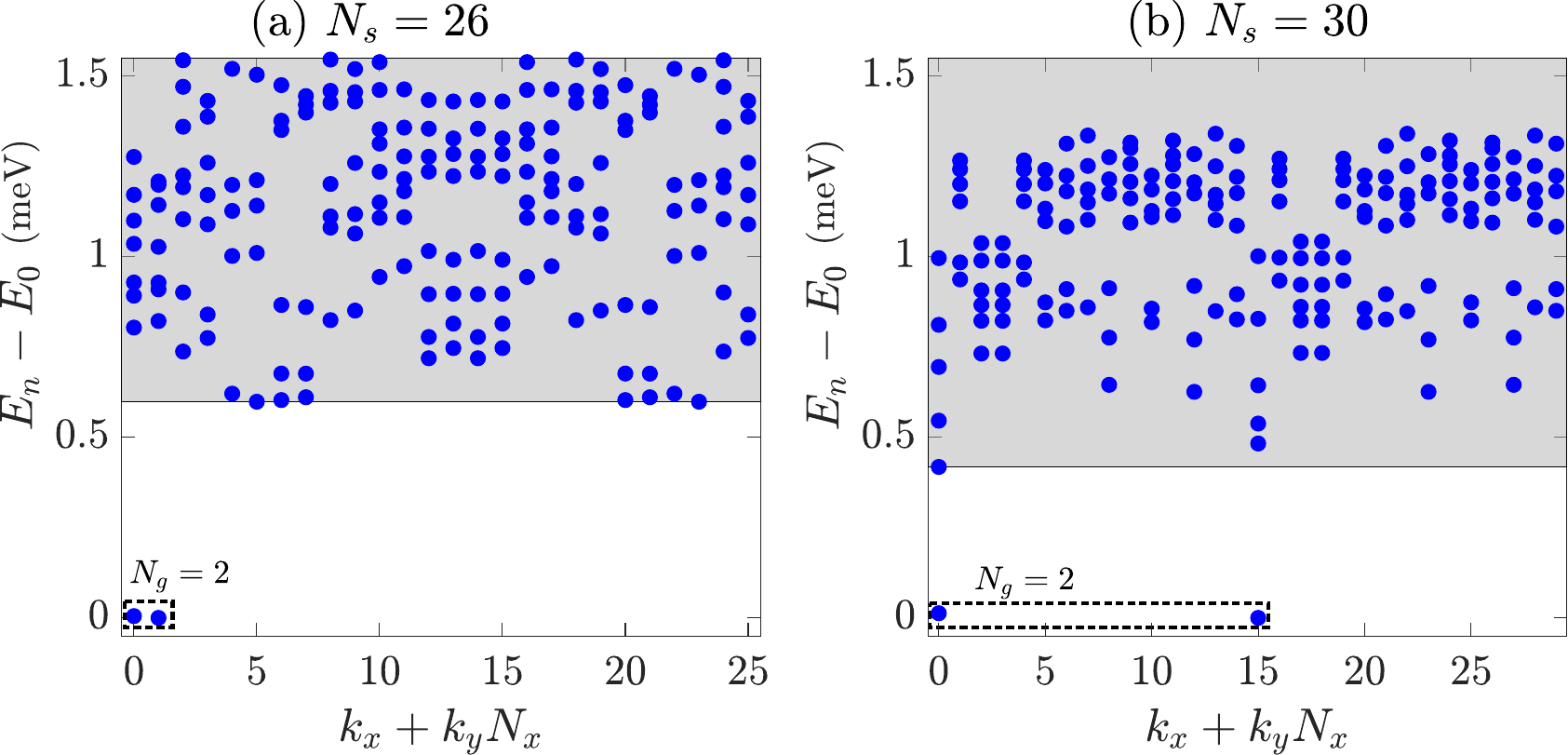}
\caption{ Energy spectra for (a) $N_s=26$ and (b) $N_s=30$ both have
odd number of holes with a ground-state degeneracy $N_g=2$. 
}\label{fig:Eng_Ns26} 
\end{figure}

    The Hartree-Fock term can be written as
        \begin{equation}\label{}  
    \begin{aligned}
       h_{\text{H}} &=  \sum_{\bm{k}_1 \bm{k}_2 }   \sum_{n m}      V_{ n m m n}(\bm{k}_1, \bm{k}_2, \bm{k}_3=\bm{k}_2, \bm{k}_4=\bm{k}_1)
      \left[ 
       \rho_{m,m}(\bm{k}_2) 
      \hat{c}^{\dagger}_{n\bm{k}_1} \hat{c}_{n\bm{k}_1}   
      +  \rho_{n,n}(\bm{k}_1) 
\hat{c}^{\dagger}_{m\bm{k}_2} 
       \hat{c}_{m\bm{k}_2} 
     \right] \\
     h_{\text{F}} &=   - \sum_{\bm{k}_1 \bm{k}_2 }   \sum_{n m}  
          V_{ m n m n}(\bm{k}_1, \bm{k}_2, \bm{k}_3=\bm{k}_1, \bm{k}_4=\bm{k}_2) 
     \left[ 
       \rho_{m,m}(\bm{k}_2) 
       \hat{c}^{\dagger}_{n\bm{k}_1} 
       \hat{c}_{n\bm{k}_1} 
      + \hat{c}^{\dagger}_{m\bm{k}_2} 
       \hat{c}_{m\bm{k}_2} 
      \rho_{n,n}(\bm{k}_1) 
     \right] 
    \end{aligned}
    \end{equation}
     where $\rho_{n,n}(\bm{k})=<\hat{c}^{\dagger}_{n\bm{k}}\hat{c}_{n\bm{k}}>$.    
    Thus we obtain the normal Hartree-Fock self-energy correction to the second moire band as
    \begin{equation}\label{}  
    \begin{aligned}
     \Sigma(\bm{k}_1) &=  2 \sum_{ \bm{k}_2 }       V_{ 2 1 1 2}(\bm{k}_1, \bm{k}_2, \bm{k}_3=\bm{k}_2, \bm{k}_4=\bm{k}_1) 
      -  2 \sum_{ \bm{k}_2 } 
    V_{  2 1 2 1}(\bm{k}_1, \bm{k}_2, \bm{k}_3=\bm{k}_1, \bm{k}_4=\bm{k}_2)   
    \end{aligned}
    \end{equation}

    In Supplementary Fig. \ref{fig:HF} (b), we plot the renormalized dispersion of the second moire band by taking into account the normal Hartree-Fock self-energy $\Sigma(\bm{k}_1)$. We find the band becomes more dispersive, i.e. the band width of non-renormalized band is around 2meV but the renormalized band width becomes larger than 9meV for $\varepsilon=5$.  

    Next we try to introduce an empirical modified Hartree-Fock self-energy function as
    \begin{equation}\label{}  
    \begin{aligned}
     \tilde{\Sigma}(\bm{k}_1) &=  2 \sum_{ \bm{k}_2 }       V_{ 2 1 1 2}(\bm{k}_1, \bm{k}_2, \bm{k}_3=\bm{k}_2, \bm{k}_4=\bm{k}_1) 
      -  2 \sum_{ \bm{k}_2 } 
    V_{ 2 1  2 1}(\bm{k}_2, \bm{k}_1, \bm{k}_3=\bm{k}_2, \bm{k}_4=\bm{k}_1)   
    \end{aligned}
    \end{equation}
    Using this empirical self-energy, the bandwidth of the renormalized band structure is relatively small, as shown in Supplementary Fig. \ref{fig:HF}(c). This empirical self-energy function is helpful to stablize the non-Abelian liquid state.   In the following sections, we will provide additional results 
    using model B with a Hamiltonian $H^B=\sum_{\mathbf k} (\varepsilon_0 (\mathbf k)+\tilde{\Sigma}(\mathbf k))\hat{c}_{2\mathbf k}^+\hat{c}_{2\mathbf k}+PH_2P$
 for $\varepsilon=5$ to show evidences of a stronger non-Abelian phase with reduced self-energy.


\section{\label{sec:energy_odd} Supplementary Note 3: Energy spectra for different system sizes
}

\begin{figure}
\includegraphics[width=0.98\linewidth,angle=0]{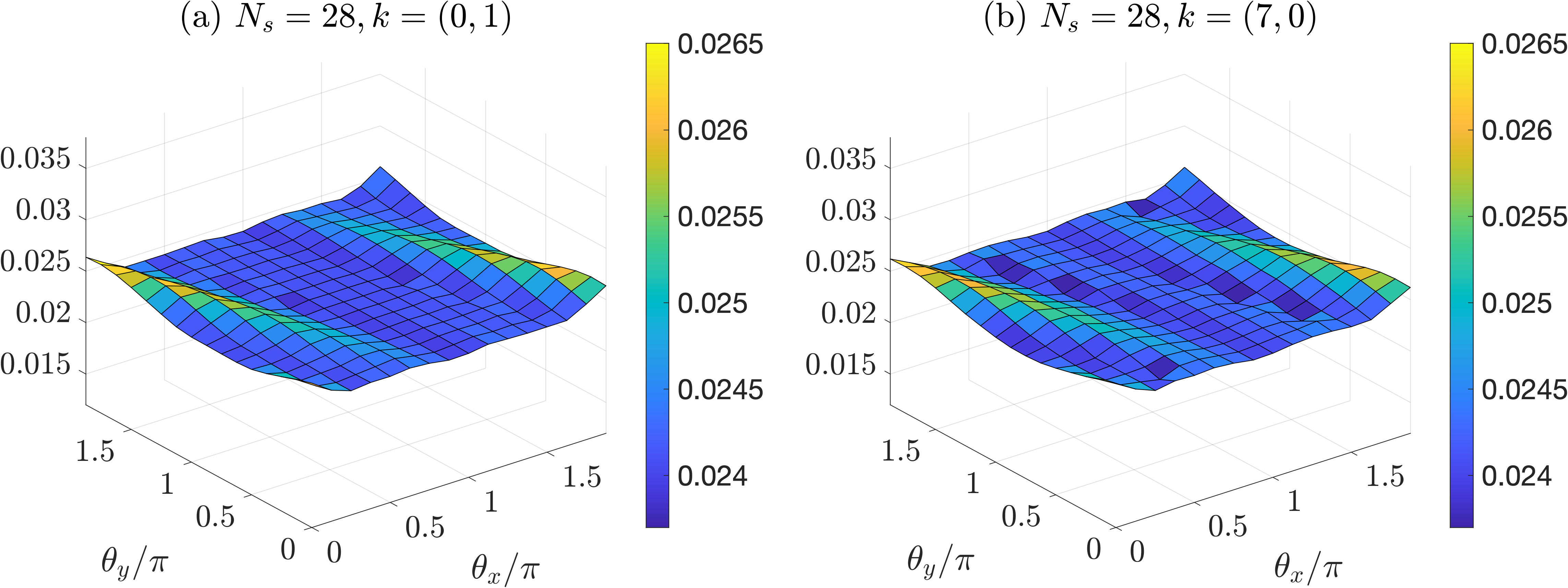}
\caption{ Berry curvature $F(\theta_x,\theta_y)\Delta\theta_x\Delta\theta_y$
   on the $N_s=28$ of Model B by discretizing the boundary phase space into $16\times16$  meshes
   in the ground state with momentum (a) $k=(0,1)$ and (b) $k=(7,0)$.
   The total Chern numbers are obtained by summing over the Berry curvature, leading to a quantized value
  $C=1/2$ for each state. } \label{fig:Berry_Ns28} 
\end{figure}

\begin{figure}
\includegraphics[width=0.98\linewidth,angle=0]{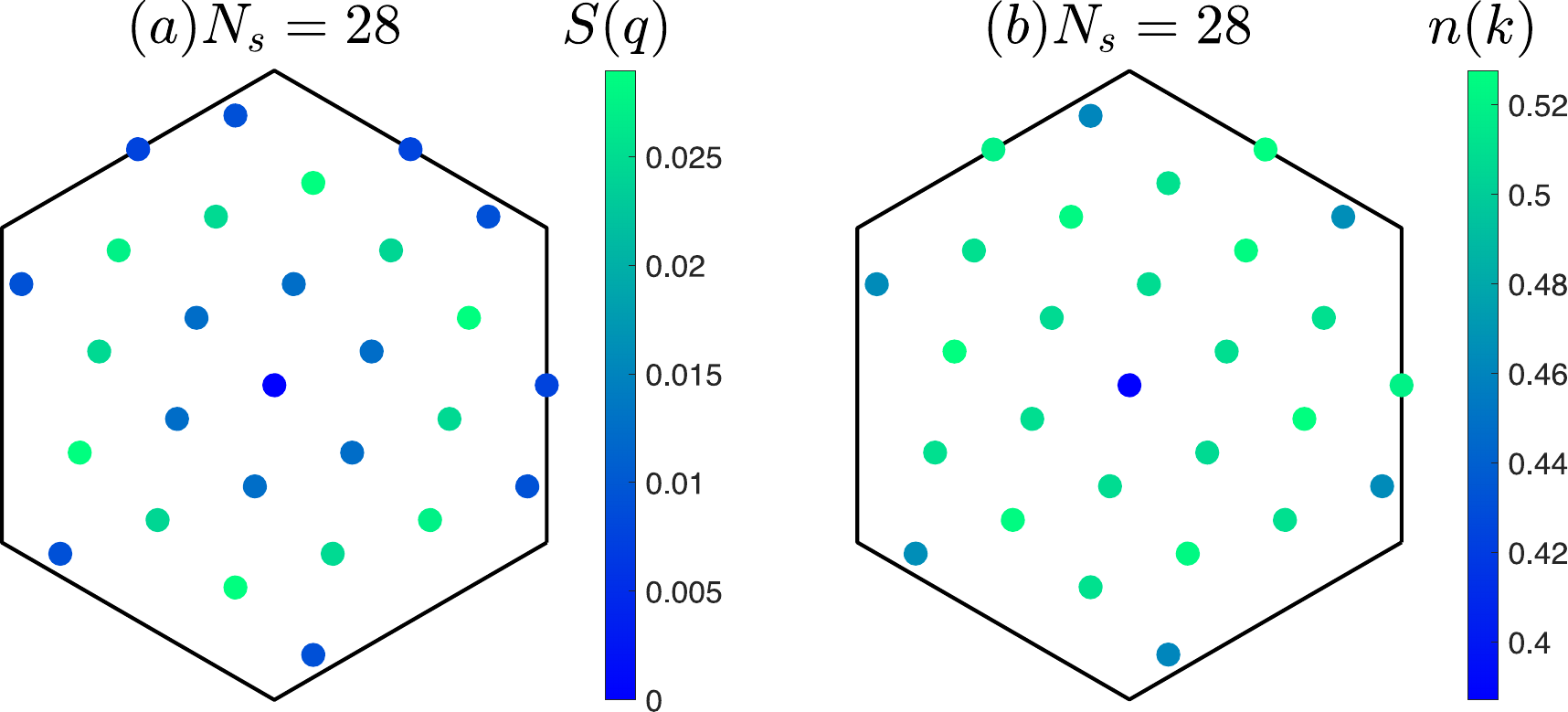}
   \caption{ (a) Density structure factor $S(\mathbf q)$ and (b) momentum distribution function $n(\mathbf k)$ of the ground state on system size $N_s=28$ with momentum $(0,0)$ of model B. } \label{fig:Sq_Ns28}
   \end{figure}

\begin{figure}[b]
   \includegraphics[width=0.6\textwidth,angle=0]{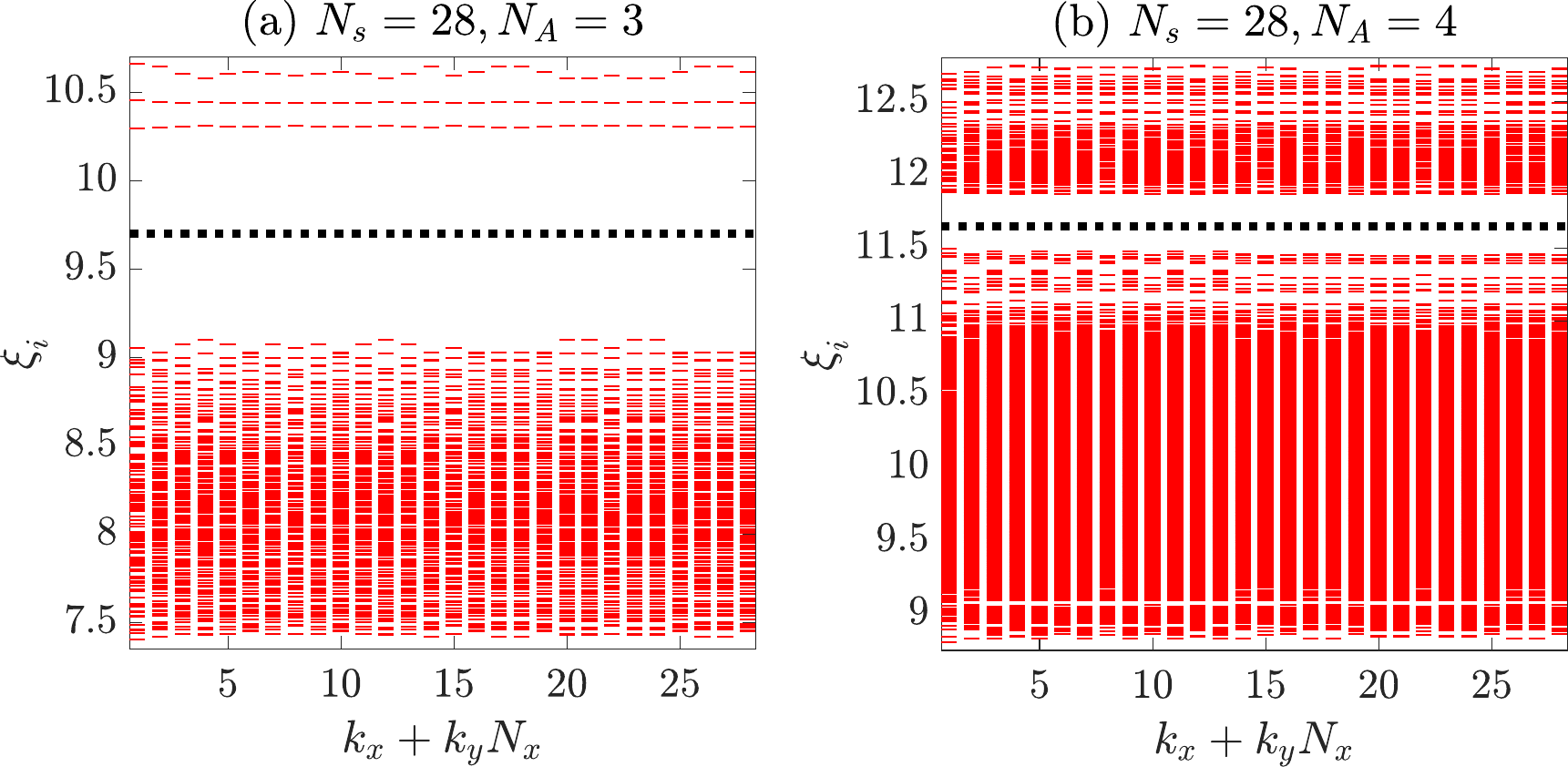}
   \caption{ The PES of Model B ($H^B$) for $N_s=28$ by dividing the system into  $N_A$ and $N_h-N_A$ particles. (a) $N_A=3$ and (b) $N_A=4$ with 3276 and 18571 levels below the entanglement gap (the dashed line) respectively. 
   The density matrix is defined as $\rho =\frac{1}{N_g} \sum_{n=1}^{N_g} | \Psi_n\rangle \langle \Psi_n|$~\cite{Sterdyniak2011}, where $|\Psi_n\rangle$ denotes the six quasi-degenerate ground states.} 
  \label{fig:PES}
\end{figure}

\begin{figure}
\includegraphics[width=0.85\linewidth,angle=0]{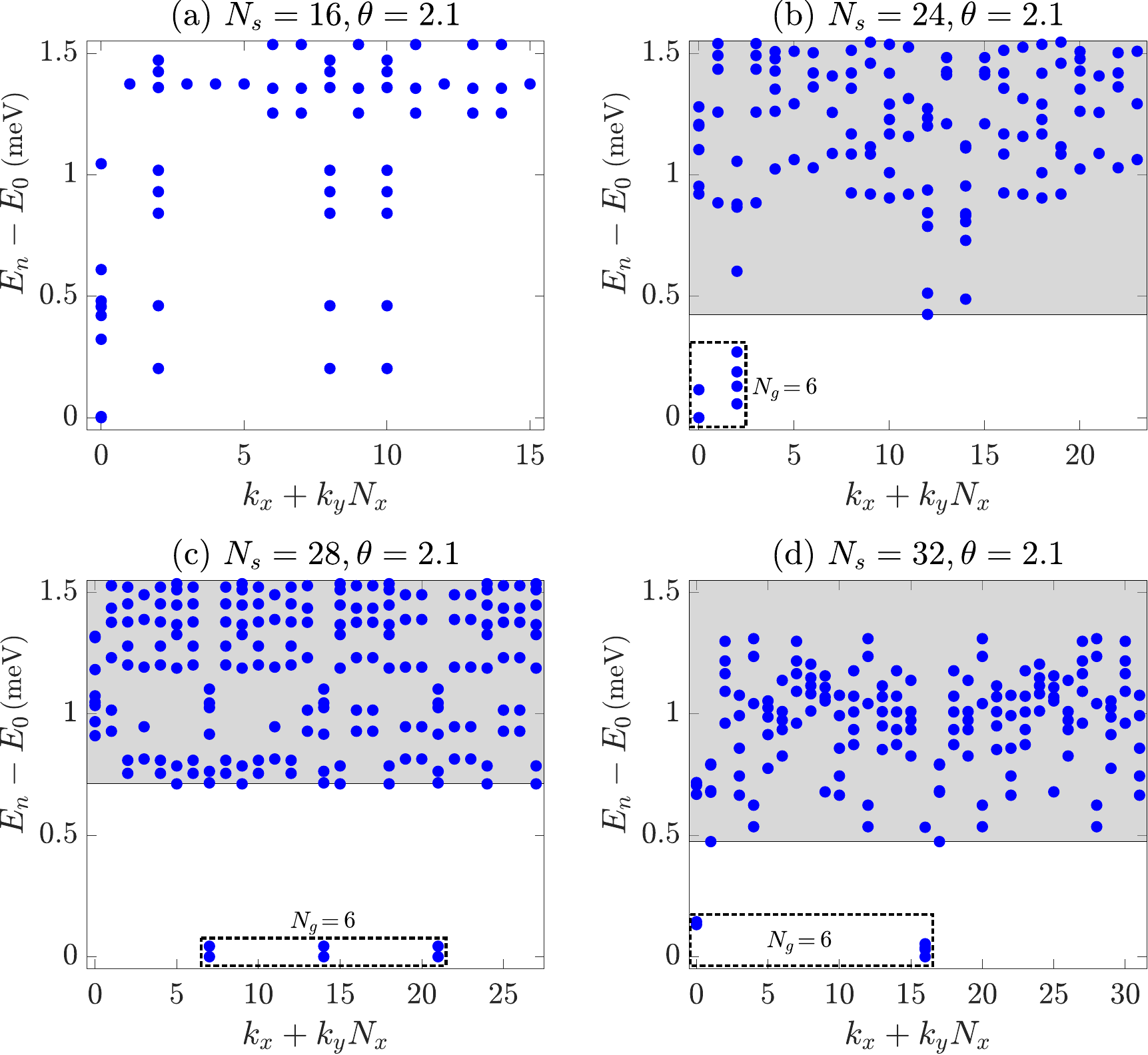}
\caption{ Energy spectra at (a) $N_s=16$, (b) $24$, (c) $28$ and (d) $32$ for $\theta=2.1$ with $V_2=-18$meV and other parameters remain the same for model B.
}\label{fig:Eng_theta21} 
\end{figure}

We consider spin polarized hole systems projected into the second lowest moir\'e energy band.
The long-range Coulomb interaction is considered with a dielectric constant
$\varepsilon=5$. We diagonalize the many-body Hamiltonian in momentum space, and label energy eigenstates with momentum quantum number ${\bf k}=k_x{\bf T}_1+k_y{\bf T}_2$ with ${\bf T}_1, {\bf T}_2$  as unit vectors of crystal momentum.
We consider a finite size system with $N_s=N_x\times N_y$ unit cells, where $N_x$, $N_y$ determine the number of discretized momentum points along $T_1$ and $T_2$ directions, respectively.

The energy spectrum for  $H^B$ of four  system sizes $N_s=16-32$ are shown in Supplementary Fig.~\ref{fig:spectrum},
where six-fold quasi-degenerate ground states (indicated by the dashed box) are found for all system sizes with moderate Coulomb interaction $\varepsilon=5$ . Significantly, the ground state manifold is well separated from the higher energy spectrum by a finite gap for all system sizes.
While our energy spectra agree with smaller system results~\cite{reddy2024nonabelian, ahn2024landau,xu2024multiple,wang2024higher}, the larger system $N_s=32$ is important for demonstrating a trend of smoothly increasing
excitation gap for $N_s=16, 24, 32$, which suggests that such a gapped phase should survive in the thermodynamic limit. 

For comparison, the energy spectra of the system sizes $N_s=26$ and $30$ are shown in Supplementary Fig. \ref{fig:Eng_Ns26} for $\theta=1.8^{o}$.   In both cases, the hole particle number $N_h=N_s/2$ 
 in the projected moir\'e band is an odd number.
 There are two ground states in different momenta sectors, in consistent with the generalized Pauli principle and the admission rule for Pfaffian (or anti-Pfaffian) state.   The even-odd effect further confirms the obtained even-denominator  FCI has non-Abelian origin.

\section{\label{sec:berry}Supplementary Note 4: Additional results for Berry curvature and Chern number 
}


To probe the topological order of the many-body state, we calculate the many-body Chern number as an integral   invariant of many-body wavefunction over twist boundary phase space.  
We discretize the boundary phase space into  $M_c\times M_c$ square meshes,
and calculate Berry phase $B_{ph}$, which is proportional to Berry curvature $F(\theta_x, \theta_y)$
as $B_{ph}=F(\theta_x,\theta_y)\Delta\theta_x\Delta\theta_y$ for   each square. 
This is done through calculating the  consecutive  wave function overlaps 
$B_{ph}(\theta_x, \theta_y)=arg(\prod_{i=1,4} \langle \Psi_i|\Psi_{i+1}\rangle)$ 
with  $|\Psi_i\rangle$ ($i$ is defined mod 4) representing four states at the four corners of the mesh square. In calculating the overlap $\langle \Psi_i|\Psi_{i+1}\rangle$, the $\Delta \theta_{x(y)}$ should be small enough so that  $|B_{ph}(\theta_x,\theta_y)|\ll \pi$ and
$|\langle \Psi_i|\Psi_{i+1}\rangle|\sim 1.0$.
We need to expand  $|\Psi_i\rangle$ into the original momentum basis for the continuum model, so that the contribution from the Berry curvature of the single particle orbitals in the moir\'e band can be taken into account.
The total Chern number is obtained  as $C=\sum \frac {B_{ph}(\theta_x, \theta_y)} {2\pi}$   over all the mesh squares in the 
$2\pi \times 2\pi$ boundary phase space.

We  show  the Berry curvature in Supplementary Fig. \ref{fig:Berry_Ns28} for larger system $N_s=28$, 
which is very uniform in the whole boundary phase space  with small fluctuations.
 By integrating the Berry curvature over the boundary phase space, 
 we obtain total  Berry phase $2\pi$ shared by two ground states. Another pair of ground states
 in the momentum $(7,1)$ sector has the same feature, which is not shown here.
 Thus  each  nearly degenerate ground state carries fractional Chern number $C=1/2$ and Hall conductance is quantized at $\sigma_{xy}=\frac 1 2 e^2/h$.
 We have confirmed the same results for other system sizes $N_s=16, 24$.

\section{\label{sec:density} Supplementary Note 5: Additional results for
density structure factor and momentum distribution function
}
 As shown in Supplementary Fig.~\ref{fig:Sq_Ns28} for a  system size $N_s=28$, the projected density structure factor
 $S({\bm q})$ for the ground state shows very weak circular-like  peaks in the middle of the moir\'e Brillouin zone, indicating  low energy density fluctuations. However,  there are no sharp peaks.
 The momentum distribution function $n(\mathbf k)$ is near uniform. Both results further confirm no instability of charge density wave.

\section{\label{sec:PES} Supplementary Note 6: Non-Abelian quasiparticle statistics from PES}
After determining the topological nature of the ground-state manifold, we further inspect the non-Abelian quasiparticle statistics through the entanglement spectroscopy with the particle-cut entanglement spectrum (PES)  \cite{Sterdyniak2011}. By dividing the whole system into $N_A$ and $N_h-N_A$ particles, we identify a clear entanglement gap separating the low-lying PES levels from higher ones (Supplementary Fig. \ref{fig:PES}) for $N_s=28$ and $N_A=3$.
 The number of PES levels below this gap exactly matches the typical counting of quasiparticle excitations resulting from the generalized Pauli principle of the non-Abelian (anti-)Pfaffian state \cite{regnault2011fractional} (at most 2 particles in 4 consecutive orbitals). 
Similar results are found for different sizes $N_s=16,24$ for B ($H^B$) model (not shown here).
Based on the correspondence between the PES and the quasiparticle excitations \cite{Sterdyniak2011}, this finding further indicates a non-Abelian nature of this half-filled state. 
Additionally, we find the gap in PES survives for $N_A=4$ for $N_s=28$ system, which shrinks 
to a smaller value  indicating there are  contributions from other low energy competing  states in consistent with results shown in main text.

\section{\label{sec:theta21} Supplementary Note 7: Energy spectrum for a different twist angle $\theta=2.1^o$}
The energy spectra of four system sizes $N_s=16$, $24$, $28$ and $32$ are 
shown in Supplementary Fig.~\ref{fig:Eng_theta21} for $\theta=2.1^o$ with a different $V_2=-18$meV while 
we keep other parameters ($V_1$, $W_1$, $W_2$, $\phi$) unchanged.  
A small change of parameter $V_2$ can make the FCI phase more robust and reduce finite size 
effect.  This is similar to tune Haldane pseudopotential in studying fractional
quantum Hall effect in Landau levels.
We can see that, for $N_s=16$, there are strong competition between the FCI
state with degeneracy at $k=(0,0)$ sector, and low energy states  at M points.
For each larger system with $N_s=24-32$,  there is a ground state manifold 
with six-fold quasi-degenerate
states (indicated by the  dashed box). In particular, the crystal momenta at which the degenerate ground states occur  precisely match those of the Pfaffian  or anti-Pfaffian state, which is consistent with the emergence of  non-Abelian state at $\theta=2.1^o$.

\end{document}